%% file: main.tex
\documentclass[nocrop]{bioinfo}
\copyrightyear{2020} \pubyear{2020}

\input{preamble}

\access{}
\appnotes{Systems biology}

\begin{document}
\firstpage{1}

\subtitle{Systems biology}

\title[Inferring Signaling Pathways with Probabilistic Programming]{Inferring Signaling Pathways with Probabilistic Programming}
\author[Merrell and Gitter]{David Merrell\,$^{\text{\sfb 1,2}}$, 
Anthony Gitter\,$^{\text{\sfb 1,2,3,}*}$ 
}
\address{$^{\text{\sf 1}}$Department of Computer Sciences, 
                          University of Wisconsin--Madison,
			  USA\\
$^{\text{\sf 2}}$Morgridge Institute for Research,
		 Madison, Wisconsin, USA\\
$^{\text{\sf 3}}$Department of Biostatistics and Medical Informatics,
                 University of Wisconsin--Madison,
                 USA.}

\corresp{$^\ast$To whom correspondence should be addressed.}

\history{}

\editor{}

\input{abstract}

\maketitle

\input{introduction}

\input{materials-methods}

\input{results}

\input{discussion}

\section*{Acknowledgements}
We thank UW-Madison's Biomedical Computing Group for generously providing compute resources;
the teams developing Snakemake and HTCondor \citep{thain-htcondor-2005} for empowering us to use those resources effectively;
the \Gen{} team \citep{cusumano-gen-2018} for designing a uniquely powerful probabilistic programming language;
and the HPN-DREAM challenge organizers for providing experimental data for our evaluation.

\section*{Funding}
This work was funded by the National Institutes of Health (award T32LM012413) and National Science Foundation (award DBI 1553206).\vspace*{-12pt}

\bibliographystyle{natbib}
\bibliography{working-bibliography/biblio}

\appendix
\beginsupplement
\input{appendices}

\end{document}

%% file: preamble.tex
\usepackage{graphicx}
\usepackage{hhline}
\usepackage{natbib}
\usepackage{xcolor}
\usepackage{outlines}
\usepackage{multirow}
\usepackage{appendix}
\usepackage{bbm}

\newcommand{\GLMNet}{\texttt{GLMNet}}
\newcommand{\Gen}{\texttt{Gen}}
\newcommand{\Ourmethodfull}{Sparse Signaling Pathway Sampling}
\newcommand{\Ourmethod}{SSPS}
\newcommand{\Funchisq}{\texttt{FunChisq}}
\newcommand{\Stan}{\texttt{Stan}}
\newcommand{\Pyro}{\texttt{Pyro}}
\newcommand{\Edward}{\texttt{Edward2}}
\newcommand{\Pymc}{\texttt{PyMC3}}

\newcommand{\primo}{\textit{(i)}}
\newcommand{\secundo}{\textit{(ii)}}
\newcommand{\tertio}{\textit{(iii)}}

\newcommand{\OOM}{\textbf{OOM}}
\newcommand{\Timeout}{\textbf{TIMEOUT}}

\newcommand{\trans}{\top}
\newcommand{\argmin}{\text{argmin}}
\newcommand{\parents}{\text{pa}}
\newcommand{\Bernoulli}{\text{Bernoulli}}
\newcommand{\Uniform}{\text{Uniform}}
\newcommand{\Neff}{N_{\text{eff}}}
\newcommand{\Nph}{N/\text{cpu-hr}}
\newcommand{\Neffph}{\Neff/\text{cpu-hr}}

\newcommand{\sizev}{|V|}
\newcommand{\rangev}{\{1\ldots|V|\}}

\newcommand{\addparent}{\texttt{add-parent}}
\newcommand{\removeparent}{\texttt{remove-parent}}
\newcommand{\swapparent}{\texttt{swap-parent}}

\newcommand{\repolink}{\href{https://github.com/gitter-lab/ssps}{https://github.com/gitter-lab/ssps}}

\let\oldpar\paragraph
\renewcommand{\paragraph}[1]{\vspace{-0.5em} \oldpar{#1}}

\newcommand{\beginsupplement}{
  \renewcommand{\thesubsection}{\Alph{subsection}}
}

%% file: abstract.tex
\abstract{ \textbf{Motivation:}
Cells regulate themselves via dizzyingly complex biochemical processes called signaling pathways. 
These are usually depicted as a network, where nodes represent proteins and edges indicate their influence on each other. 
In order to understand diseases and therapies at the cellular level, it is crucial to have an accurate understanding of the signaling pathways at work. 
Since signaling pathways can be modified by disease, the ability to infer signaling pathways from condition- or patient-specific data is highly valuable. \\ 
A variety of techniques exist for inferring signaling pathways. 
We build on past works that formulate signaling pathway inference as a Dynamic Bayesian Network structure estimation problem on phosphoproteomic time course data. 
We take a Bayesian approach, using Markov Chain Monte Carlo to estimate a posterior distribution over possible Dynamic Bayesian Network structures. 
Our primary contributions are \primo{} a novel proposal distribution that efficiently samples sparse graphs and \secundo{} the relaxation of common restrictive modeling assumptions.\\
\textbf{Results:}
We implement our method, named \Ourmethodfull{}, in Julia using the Gen probabilistic programming language. 
Probabilistic programming is a powerful methodology for building statistical models. 
The resulting code is modular, extensible, and legible. 
The Gen language, in particular, allows us to customize our inference procedure for biological graphs and ensure efficient sampling.\\
We evaluate our algorithm on simulated data and the HPN-DREAM pathway reconstruction challenge, comparing our performance against a variety of baseline methods.
Our results demonstrate the vast potential for probabilistic programming, and Gen specifically, for biological network inference.\\
\textbf{Availability:} Find the full codebase at \repolink{} \\
\textbf{Contact:} gitter@biostat.wisc.edu
}

%% file: introduction.tex
\section{Introduction}
\label{sec:intro}

Signaling pathways enable cells to process information rapidly in response to external environmental changes or intracellular cues.
One of the core signaling mechanisms is protein phosphorylation.
Kinases add phosphate groups to substrate proteins and phosphatases remove them.
These changes in phosphorylation state can act as switches, controlling proteins' activity and function.
A protein's phosphorylation status affects its localization, stability, and interaction partners \citep{2014-newman-toward}.
Ultimately, phosphorylation changes regulate important biological processes such as transcription and cell growth, death, and differentiation \citep{2009-hunter-tyrosine, kholodenko-signalling-2010}.

Pathway databases characterize the signaling relationships among groups of proteins but are not tailored to individual biological contexts.
Even for well-studied pathways such as epidermal growth factor receptor-mediated signaling, the proteins significantly phosphorylated during a biological response can differ greatly from those in the curated pathway \citep{2018-koksal-tps}.
The discrepancy can be due to context-specific signaling \citep{2017-hill-context}, cell type-specific protein abundances, or signaling rewiring in disease \citep{2007-pawson-oncogenic}.
Therefore, there is a need to learn context-specific signaling pathway representations from observed phosphorylation changes.
In the clinical setting, patient-specific signaling pathway representations may eventually be able to guide therapeutic decisions \citep{2016-drake-phosphoproteome, halasz-integrating-2016, eduati-patientspecific-2020}.

Diverse classes of techniques have been developed to model and infer signaling pathways \citep{kholodenko-flow-2012}.
They take approaches including Granger causality \citep{shojaie-granger-2010, carlin-prophetic-2017}, information theory \citep{cheong-information-2011, krishnaswamy-conditional-2014}, logic models \citep{eker-pathway-2002, guziolowski-exhaustively-2013, gjerga-converting-2020}, differential equations \citep{schoeberl-computational-2002, 2013-molinelli-perturbation, henriques-datadriven-2017}, non-parametric statistical tests \citep{zhang-funchisq-2013}, and probabilistic graphical models \citep{sachs-causal-2005} among others.
Some signaling pathway reconstruction algorithms take advantage of perturbations such as receptor stimulation or kinase inhibition.
Although perturbing individual pathway members can causally link them to downstream phosphorylation changes, characterizing a complex pathway can require a large number of perturbation experiments.
Inferring pathway structure from temporal phosphorylation data presents an attractive alternative.
A single time series phosphorylation dataset can reveal important dynamics without perturbing individual pathway members.
For instance, a kinase cannot phosphorylate substrates before it is activated.

An alternative approach to pathway reconstruction selects a context-specific subnetwork from a general background network.
These algorithms can use phosphorylation data to assign scores to protein nodes in a protein-protein interaction network.
They then select edges that connect the high-scoring nodes, generating a subnetwork that may explain how the induced phosphorylation changes arise from the source of stimulation.
Extensions accommodate temporal scores on the nodes \citep{patil-linking-2013, budak-reconstruction-2015, 2018-koksal-tps, norman-ststeiner-2019}.

Our present work builds on past techniques that formulate signaling pathway inference as a Dynamic Bayesian Network (DBN) structure estimation problem.
This family of techniques relies on two core ideas:
\primo{} we can use a DBN to model phosphorylation time series data; and
\secundo{} the DBN's structure translates directly to a directed graph representing the signaling pathway.
Rather than identifying a single DBN that best fits the data, these techniques take a Bayesian approach---they yield a \emph{posterior distribution} over possible DBN structures.
Some techniques use Markov Chain Monte Carlo (MCMC) to sample from the posterior \citep{2007-werhli-bayesian, 2010-gregorczyk}.
Others use exact, enumerative inference to compute posterior probabilities \citep{hill-bayesian-2012, 2014-oates-joint, 2015-spencer-interventional}. 

We present a new Bayesian DBN-based technique, \Ourmethodfull{} (\Ourmethod{}).
It improves on past MCMC methods by using a novel proposal distribution specially tailored for the large, sparse graphs prevalent in biological applications.
Furthermore, \Ourmethod{} makes weaker modeling assumptions than other DBN approaches.
As a result, \Ourmethod{} scales to larger problem sizes and yields superior predictions in comparison to other DBN techniques.

We implement \Ourmethod{} using the \Gen{} probabilistic programming language (PPL).
Probabilistic programming is a powerful methodology for building statistical models.
It enables the programmer to build models in a legible, modular, reusable fashion.
This flexibility was important for prototyping and developing the current form of \Ourmethod{} and readily supports future improvements or extensions.

%% file: materials-methods.tex
\section{Materials and methods}
\label{sec:method}

\subsection{Model formulation}
\label{sec:model}

\Ourmethod{} makes specific modeling assumptions.
We start with the DBN model of \cite{hill-bayesian-2012}, relax some assumptions, and modify it in other ways to be better-suited for MCMC inference.

\paragraph{Preliminary definitions.}
We first define some notation for clarity's sake.
Let $G$ denote a \emph{directed graph} with vertices $V$ and edges $E(G)$.
Graph $G$ will represent a signaling pathway, with vertices $V$ corresponding to proteins and edges $E(G)$ indicating their influence relationships.
We use $\parents_G(i)$ to denote the \emph{parents} of vertex $i$ in $G$.
 
Let $X$ denote our time series data, consisting of $\sizev$ variables measured at $T$ timepoints.
$X$ is a $T {\times} \sizev$ matrix where the $j$th column corresponds to the $j$th variable and the $j$th graph vertex.
As a convenient shorthand, let $X_+$ denote the \emph{latest} $T{-}1$ timepoints in $X$,
and let $X_-$ denote the \emph{earliest} $T{-}1$ timepoints in $X$.
Lastly, define $B_j \equiv X_{-,\parents_G(j)}$.
In other words, $B_j$ contains the values of variable $j$'s parents at the $T{-}1$ earliest timepoints.
In general, $B_j$ may also include columns of nonlinear interactions between the parents. 
We will only include linear terms, unless stated otherwise.

\paragraph{Model derivation.}
In our setting, we aim to infer $G$ from $X$.
In particular, Bayesian approaches seek a \emph{posterior distribution} $P(G|X)$ over possible graphs.
From Bayes's rule we know 
    $P(G|X) \propto P(X|G) \cdot P(G).$
That is, a Bayesian model is fully specified by its choice of \emph{prior distribution} $P(G)$
and \emph{likelihood function} $P(X|G)$.

We derive our model from the one used by \cite{hill-bayesian-2012}.
They choose a prior distribution of the form
\begin{equation}
P(G~|~G^\prime, \lambda) \propto \exp\left( -\lambda | E(G) \setminus E(G^\prime)| \right) \label{eq:oldprior}
\end{equation}
parameterized by a \emph{reference graph} $G^\prime$ and \emph{inverse temperature} $\lambda$.
This prior gives uniform probability to all subgraphs of $G^\prime$ and ``penalizes'' edges not contained in $E(G^\prime)$.
$\lambda$ controls the ``importance'' given to the reference graph.

\begin{figure}[t!]
    \centering
    \includegraphics[scale=0.75]{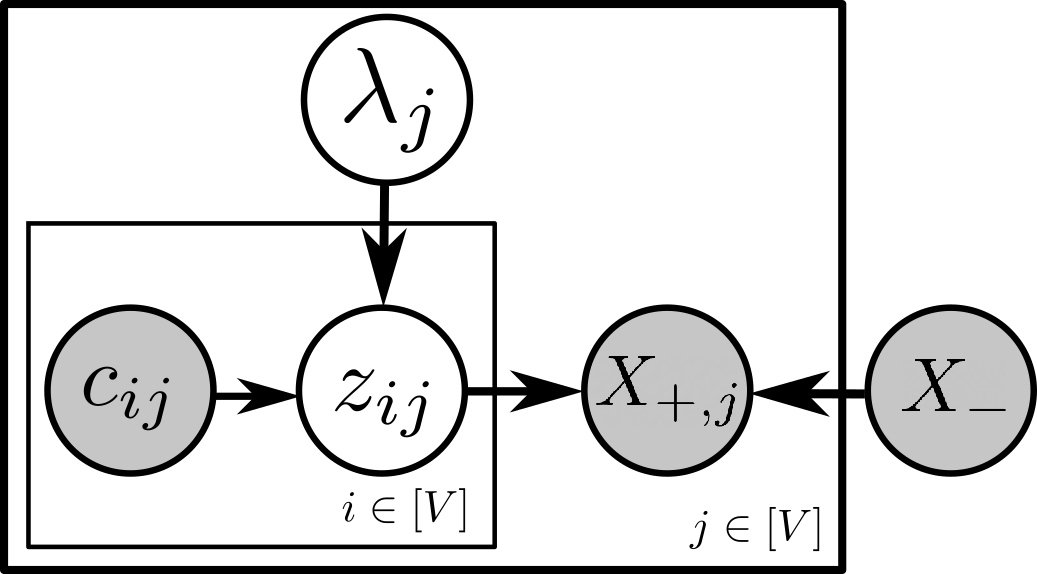}
    \resizebox{1.0\linewidth}{!}{
    \begin{minipage}{\linewidth}
    \begin{align*}
        \lambda_j &\sim \Uniform(\lambda_{\text{min}}, \lambda_{\text{max}}) &\forall j \in \rangev \\[-3pt]
        z_{ij}~|~c_{ij}, \lambda_j &\sim \Bernoulli\left(\frac{e^{-\lambda_j}}{e^{-c_{ij} \lambda_j} + e^{-\lambda_j}}\right) &\forall i, j \in \rangev \\[4pt]
        \sigma_j^2 &\propto \frac{1}{\sigma_j^2} & \forall j \in \rangev \\[-3pt]
        \beta_j~|~\sigma_j^2 &\sim \mathcal{N}\left(0, T\sigma_j^2 (B_j^\trans B_j)^{-1} \right) & \forall j\in \rangev\\[4pt]
        X_{+,j}~|~B_j, \beta_j, \sigma_j^2 &\sim \mathcal{N}\left( B_j \beta_j, \sigma_j^2 I \right) & \forall j\in \rangev 
    \end{align*}
    \end{minipage}
}
    \caption{Our generative model. (top) Plate notation. DBN parameters $\beta_j$ and $\sigma_j^2$ have been marginalized out. 
    (bottom) Full probabilistic specification.
    We usually set $\lambda_\text{min} \simeq 3$  and $\lambda_\text{max} {=} 15$. 
    If $\lambda_\text{min}{>}0$ is too small, Markov chains will occasionally be initialized with very large numbers of edges, causing computational issues. 
    The method is insensitive to $\lambda_\text{max}$ as long as it's sufficiently large. 
    Notice the improper prior $1/\sigma_j^2$. 
    In this specification $B_j$ denotes $X_{-, \parents_Z(j)}$; that is, the parents of vertex $j$ depend on edge existence variables $Z$.
    }
    \label{fig:new-model}
\end{figure}

\citeauthor{hill-bayesian-2012} choose a Gaussian DBN for their likelihood function.
Intuitively, they assume linear relationships between variables and their parents:
\vspace{-1.1em}
$$ 
X_{+,j} \sim \mathcal{N}(B_j \beta_j, \sigma_j^2) \hspace{4em} \forall j \in \rangev.
$$
A suitable prior over the regression coefficients $\beta_j$ and noise parameters $\sigma_j^2$ (Figure \ref{fig:new-model}) allows us to integrate them out, yielding this \emph{marginal likelihood function}:
\begin{equation} P(X|G) \propto \prod\limits_{j=1}^{\sizev} T^{-\frac{|\parents_G(j)|}{2}} \left(X_{+,j}^{\trans} X_{+,j} - \frac{T {-} 1}{T} X_{+,j}^\trans (B_j \hat{\beta}_{ols}) \right)^{-\frac{T-1}{2}} \label{eq:marglik}
\end{equation}                
where $\hat{\beta}_{ols} = (B_j^\trans B_j)^{-1} B_j^\trans X_{+,j}$ is the ordinary least squares estimate of $\beta_j$.
For notational simplicity, Equation \ref{eq:marglik} assumes we have a single time course of length $T$. 
In general, there may be multiple time course replicates with differing lengths.
The marginal likelihood generalizes to that case in a straightforward way.

In \Ourmethod{} we use the same marginal likelihood function (Equation \ref{eq:marglik}), but a different prior distribution $P(G)$.
We obtain our prior distribution by decomposing Equation \ref{eq:oldprior} into a product of independent Bernoulli trials over graph edges.
This decomposition in turn allows us to make some useful generalizations.
Define \emph{edge existence variables} $z_{ij} \equiv \mathbbm{1}[(i,j) \in E(G)]$.
Let $Z$ be the $\sizev {\times} \sizev$ matrix of all $z_{ij}$.
Then we can rewrite Equation \ref{eq:oldprior} as follows:
$$
P(G|G^\prime, \lambda) ~\equiv~ P(Z|G^\prime, \lambda) ~\propto~ \prod_{(i,j) \notin E(G^\prime)}\!\!\! e^{-z_{ij} \lambda} 
$$
$$
= \prod_{(i,j) \in E(G^\prime)}\!\!\! \left(\frac{1}{2}\right)^{z_{ij}}\!\left(\frac{1}{2}\right)^{1 {-} z_{ij}}\!\!\!\!\! \prod_{(i,j) \notin E(G^\prime)} \left( \frac{e^{-\lambda}}{1 {+} e^{-\lambda}} \right)^{z_{ij}}\! \left( \frac{1}{1 {+} e^{-\lambda}} \right)^{1 {-} z_{ij}}
$$
where the last line is a true equality---it gives a normalized probability measure.
We see that the original prior is simply a product of Bernoulli variables parameterized by a shared inverse temperature, $\lambda$.
See Appendix \ref{sec:append-model} for a more detailed derivation.

Rewriting the prior in this form opens the door to generalizations.
First, we address a shortcoming in the way reference graph $G^\prime$ expresses prior knowledge.
The original prior assigns equal probability to every edge of $G^\prime$.
However, in practice we may have differing levels of prior confidence in the edges.
We address this by allowing a real-valued prior confidence $c_{ij}$ for each edge:
\begin{equation}
    P(Z|C,\lambda) =  \prod_{(i,j)} \left( \frac{e^{-\lambda}}{e^{-c_{ij}\lambda} {+} e^{-\lambda}} \right)^{z_{ij}}\! \left( \frac{e^{-c_{ij}\lambda}}{e^{-c_{ij}\lambda} {+} e^{-\lambda}} \right)^{1 {-} z_{ij}} \label{eq:prior-conts}
\end{equation}
where $C$ is the matrix of all prior confidences $c_{ij}$, replacing $G^\prime$.
Notice that if every $c_{ij} {\in} \{0,1\}$, then Equation \ref{eq:prior-conts} is equivalent to the original prior.
In effect, Equation \ref{eq:prior-conts} \emph{interpolates} the original prior, permitting a continuum of confidences on the interval $[0,1]$.

We make one additional change to the prior by replacing the shared $\lambda$ inverse temperature variable with a collection of variables, $\Lambda = \{ \lambda_j ~|~ j = 1, {\ldots} ,|V| \}$, 
one for each vertex of the graph.
Recall that the original $\lambda$ variable determined the importance of the reference graph.
In the new formulation, each $\lambda_j$ controls the importance of the prior knowledge for vertex $j$ and its parents:
\begin{equation}
    P(Z|C,\Lambda) =  \prod_{(i,j)} \left( \frac{e^{-\lambda_j}}{e^{-c_{ij}\lambda_j} {+} e^{-\lambda_j}} \right)^{z_{ij}}\! \left( \frac{e^{-c_{ij}\lambda_j}}{e^{-c_{ij}\lambda_j} {+} e^{-\lambda_j}} \right)^{1 {-} z_{ij}} \label{eq:new-prior}
\end{equation}
We introduced $\Lambda$ primarily to help MCMC converge more efficiently.
Experiments with the shared $\lambda$ revealed a multimodal posterior that tended to trap $\lambda$ in high or low values.
The introduction of vertex-specific $\lambda_j$ variables yielded faster convergence with weaker modeling assumptions---an improvement in both respects.

We implicitly relax the model assumptions further via our inference procedure.
For sake of tractability, the original exact method of \cite{hill-bayesian-2012} imposes a hard constraint on the in-degree of each vertex.
In contrast, we use a MCMC inference strategy with no in-degree constraints.

In summary, our model departs from that of \cite{hill-bayesian-2012} in three important respects.
It permits real-valued prior confidences $C$, 
introduces vertex-specific inverse temperature variables $\Lambda$,
and places no constraints on vertices' in-degrees.
See the full model in Figure \ref{fig:new-model} and Appendix \ref{sec:append-model} for additional details.

\subsection{Inference procedure}
\label{sec:inference}

Our method uses MCMC to infer posterior edge existence probabilities.
As described in Section \ref{sec:model}, our model contains two classes of unobserved random variables: \primo{} the edge existence variables $Z$ and \secundo{} the inverse temperature variables $\Lambda$.
For each step of MCMC, we loop through these variables and update them in a Metropolis-Hastings fashion.

\paragraph{Main loop.}
At a high level, our MCMC procedure consists of a loop over the graph vertices, $V$.
For each vertex $j$, we update its inverse temperature variable $\lambda_j$
and then update its \emph{parent set} $\parents_G(j)$.
All of these updates are Metropolis-Hastings steps; the proposal distributions are described below.
Each completion of this loop yields one iteration of the Markov chain.

\paragraph{Proposal distributions.}
For the inverse temperature variables we use a symmetric Gaussian proposal:
$\lambda_j^\prime \sim \mathcal{N}(\lambda_j, \xi^2)$. 
In practice the method is insensitive to $\xi$; we typically set $\xi{=}3$.

The parent set proposal distribution is more complicated.
There are two principles at work when we design a graph proposal distribution:
\primo{} the proposal should efficiently traverse the space of directed graphs, and
\secundo{} it should favor graphs with higher posterior probability.
The most widely used graph proposal distribution selects a \emph{neighboring} graph uniformly from the set of possible ``add-edge,'' ``remove-edge,'' and ``reverse-edge'' actions \citep{2007-werhli-bayesian, 2010-gregorczyk}.
We'll refer to this traditional proposal distribution as the \emph{uniform graph proposal}. 
In our setting, we expect sparse graphs to be much more probable than dense ones---notice how the marginal likelihood function (Equation \ref{eq:marglik}) strongly penalizes $|\parents_G(j)|$.
However, the uniform graph proposal exhibits a preference toward \emph{dense graphs}.
It proposes ``add-edge'' actions too often.
This motivates us to design a new proposal distribution tailored for sparse graphs---one that operates on our sparse \emph{parent set} graph representation.

For a given graph vertex $j \in V$, the parent set proposal distribution updates $\parents_G(j)$ by choosing from the following actions:
\begin{itemize}
    \item \addparent{}. Select one of vertex $j$'s non-parents uniformly at random, and add it to $\parents_G(j)$.
    \item \removeparent{}. Select one of vertex $j$'s parents uniformly at random, and remove it from $\parents_G(j)$.
    \item \swapparent{}. A simultaneous application of \addparent{} and \removeparent{}. 
          Perhaps surprisingly, this action is not made redundant by the other two.
          It plays an important role by yielding updates that maintain the size of the parent set.
          Because the marginal likelihood (Equation \ref{eq:marglik}) changes steeply with $|\parents_G(j)|$,
          Metropolis-Hastings acceptance probabilities will be higher for actions that keep $|\parents_G(j)|$ constant.
\end{itemize}
These three actions are sufficient to explore the space of directed graphs, but we need another mechanism to bias the exploration toward \emph{sparse} graphs.
We introduce this preference via the \emph{probability} assigned to each action.
Intuitively, we craft the action probabilities so that when $|\parents_G(j)|$ is too small, \addparent{} moves are most probable.
When $|\parents_G(j)|$ is too big, \removeparent{} moves are most probable.
When $|\parents_G(j)|$ is about right, all moves are equally probable. 

\begin{figure}
    \includegraphics[scale=0.65]{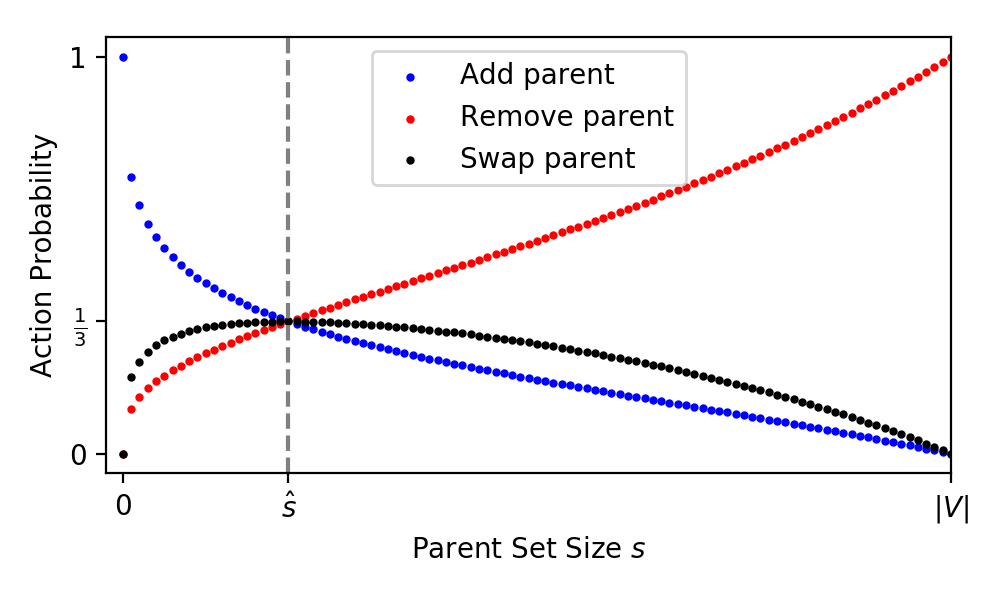}
    \caption{Action probabilities as a function of parent set size.
             The reference size $\hat{s}$ is determined from prior knowledge.
             It approximates the size of a ``typical'' parent set.
             When $s {<} \hat{s} $, \addparent{} is most probable;
             when $s {>} \hat{s} $, \removeparent{} is most probable;
             and when $s {=} \hat{s} $,  all actions have equal probability.
             }
    \label{fig:action-probs}
\end{figure}

We formulate the action probabilities for vertex $j$ as follows.
As a shorthand, let $s_j = |\parents_G(j)|$ and define the \emph{reference size} $\hat{s}_j = \sum_{i=1}^{\sizev} c_{ij}$. 
That is, $\hat{s}_j$ uses the prior edge confidences $C$ to estimate an appropriate reference size for the parent set. 
Then, the action probabilities are
\vspace{-1.1em}
$$
\begin{aligned}
    p(\addparent | s_j, \hat{s}_j) & \propto 1 - \left(\frac{s_j}{\sizev}\right)^{\gamma(\hat{s}_j)} \\ 
    p(\removeparent | s_j, \hat{s}_j) & \propto \left(\frac{s_j}{\sizev} \right)^{\gamma(\hat{s}_j)}\\ 
    p(\swapparent | s_j, \hat{s}_j) & \propto 2\left(\frac{s_j}{\sizev} \right)^{\gamma(\hat{s}_j)} \cdot \left(1 - \left(\frac{s_j}{\sizev}\right)^{\gamma(\hat{s}_j)}\right)
\end{aligned}
$$
where $\gamma(\hat{s}_j) =  1 / \log_2(\sizev / \hat{s}_j)$.
We use these functional forms only because they have certain useful properties:
\primo{} when $s_j{=}0$, the probability of \addparent{} is 1;
\secundo{} when $s_j{=}\sizev$, the probability of \removeparent{} is 1;
and \tertio{} when $s_j{=}\hat{s}_j$, all actions have equal probability (Figure \ref{fig:action-probs}).
Beyond that, these probabilities have no particular justification.
We provide additional information about the parent set proposal in Appendix \ref{sec:append-proposal}.

Recall that Metropolis-Hastings requires us to compute the reverse transition probability for any proposal we make.
This could pose a challenge given our relatively complicated parent set proposal distribution.
However, \Gen{} provides a helpful interface for computing reverse probabilities.
The user can provide an \emph{involution} function that returns the reverse of a given action.
\Gen{} then manages the reverse probabilities without further intervention.
This makes it relatively easy to implement Metropolis-Hastings updates with unusual proposal distributions.

\paragraph{Termination, convergence, and inference.}
We follow the basic MCMC protocols described by \cite{gelman-text-2014}. 
This entails running multiple (i.e., 4) Markov chains and discarding the first half of each chain as burnin.
In all of our analyses, we terminate each Markov chain when it either \primo{} reaches a length of 100,000 iterations or \secundo{} the execution time exceeds 12 hours.
These termination conditions are arbitrary but emulate a real-world setting where it may be acceptable to let the method run overnight.

Upon termination, we assess convergence with two diagnostics: Potential Scale Reduction Factor (PSRF) and effective number of samples ($\Neff$).
PSRF identifies cases where the Markov chains fail to mix or achieve stationarity.
$\Neff$ provides a sense of ``sample size'' for our inferred quantities.
It adjusts the number of MCMC samples by accounting for autocorrelation in each chain.
For our purposes, we say a quantity has \emph{failed to converge} if its PSRF $\geq {1.01}$ or $\Neff {<} 10$.
Note that satisfying these criteria does not guarantee convergence.
However, failure to satisfy them is a reliable flag for non-convergence.

Assuming a quantity hasn't failed to converge, we estimate it by simply taking its sample mean from all samples remaining after burnin.
In our setting we are primarily interested in \emph{edge existence} probabilities; 
i.e., we compute the fraction of samples containing each edge.

\subsection{Probabilistic programming implementation}
\label{sec:pp}
We implemented \Ourmethod{} using the \Gen{} PPL.
We briefly describe the probabilistic programming methodology
and its advantages in our setting.

\paragraph{Probabilistic programming.}
Probabilistic programming is a methodology for building statistical models.
It's based on the idea that statistical models are \emph{generative processes}---sequences of operations on random variables.
In probabilistic programming, we express the generative process as a program written in a PPL. 
This program is then compiled to produce a log-probability function, which can be used in inference tasks.
Probabilistic programming systems typically provide a set of generic inference methods for performing those tasks---e.g., MCMC or Variational Bayes.    

Compare this with a more traditional approach, where the user must
\primo{} derive and implement the log-probability function and 
\secundo{} implement an inference method that operates on the log-probability function.
This process of manual derivation and implementation is error-prone and requires a high degree of expertise from the user.
In contrast, probabilistic programming only requires the user to express their model in a PPL.
The probabilistic programming system manages other details.

Probabilistic programming also tends to promote good software engineering principles such as abstraction, modularity, and legibility.
Most PPLs organize code into functions, which can be reused by multiple statistical models.

\paragraph{Probabilistic programming languages.}
Several PPLs have emerged in recent years.
Examples include 
\Stan{} \citep{carpenter-stan-2017},
\Edward{} \citep{dillon-tfp-2017}, 
\Pyro{} \citep{bingham-pyro-2018},
\Pymc{} \citep{salvatier-pymc3-2016}, and 
\Gen{} \citep{cusumano-gen-2018}.
PPLs differ in how they balance \emph{expressive power} and \emph{ease of use}. 
For example, \Stan{} makes it easy to build hierarchical statistical models with continuous variables but caters poorly to other model classes.
At the other extreme, \Gen{} can readily express a large class of models---discrete and continuous variables with complex relationships---but requires the user to design a customized inference procedure.

\paragraph{Implementation in \Gen{}.}
We use the \Gen{} PPL precisely for its expressive power and customizable inference.
While implementing \Ourmethod{}, the customizability of \Gen{} allowed us to begin with simple prototypes and then make successive improvements.
For example, our model initially used a dense \emph{adjacency matrix} representation for $G$, but subsequent optimizations led us to use a sparse \emph{parent set} representation instead.
Similarly, our MCMC method started with a na\"ive ``add or remove edge'' proposal distribution; we arrived at our sparse proposal distribution (Section \ref{sec:inference}) after multiple refinements. 
Other PPLs do not allow this level of control (Table \ref{tab:ppls}).

\begin{table}
\processtable{} 
    {\begin{tabular}{|@{}m{13mm}|m{18mm}|m{25mm}|m{22mm}|@{}} \hline \ PPL & Host language & Primary \newline model class & Primary inference method \\ \hhline{|=|=|=|=|}
\ \Stan{} & custom language & hierarchical, cont's vars & Black-box HMC\\ \hline
\ \Edward{} & Python/ TensorFlow & ``deep'', cont's vars & Black-box variational\\ \hline
\ \Pymc{} & Python/Theano & ``deep'', cont's vars & Black-box HMC\\ \hline
\ \Pyro{} & Python/PyTorch & ``deep'', cont's vars & Black-box variational\\ \hline
\ \Gen{} & Julia & discrete and cont's vars; highly flexible & Customizable MCMC\\\hline
	\end{tabular}}{}
    \caption{A coarse comparison of some noteworthy PPLs. 
        \Gen{} provides expressiveness but requires the user to implement an inference program for their model.
		Cont's vars: continuous variables; HMC: Hamiltonian Monte Carlo.}
    \label{tab:ppls}
\end{table}

\subsection{Simulation study evaluation}
\label{sec:simulation}

We use a simulation study to answer important questions about \Ourmethod{}:
How does its computational expense grow with problem size?
Is it able to correctly identify true edges?
What is its sensitivity to errors in the prior knowledge?
Simulations allow us to answer these questions in a controlled setting where we have access to ground truth.

\paragraph{Data simulation process.}    
We generate each simulated dataset as follows:
\begin{enumerate}
    \item Sample a random adjacency matrix $A \in \{0,1\}^{\sizev {\times} \sizev}$, where each entry is the outcome of a $\Bernoulli(p)$ trial.
        $A$ specifies the \emph{structure} of a DBN.
        We choose $p{=}5/\sizev$ so that each vertex has an average of 5 parents.
				This approximates the sparsity we might see in signaling pathways.
        We denote the size of the original edge set as $|E_0|$.
    \item Let the weights $\beta$ for this DBN be drawn from a normal distribution $\mathcal{N}(0,1/\sqrt{\sizev})$.
        We noticed empirically that the $1/\sqrt{\sizev}$ scale prevented the simulated time series from diverging to infinity.
    \item Use the DBN defined by $A, \beta$ to simulate $M$ time courses of length $T$.
        We imitate the real datasets in Section \ref{sec:dream} by generating $M{=}4$ time courses, each of length $T{=}8$.
    \item Corrupt the adjacency matrix $A$ in two steps: \primo{} remove $r \cdot |E_0|$ of the edges from $A$;
        \secundo{} add $a \cdot |E_0|$ spurious edges to the adjacency matrix.
        This corrupted graph simulates the \emph{imperfect prior knowledge} encountered in reality.
        The parameters $r$ and $a$ control the ``false negatives'' and ``false positives'' in the prior knowledge, respectively.
\end{enumerate}
We use a range of values for parameters $\sizev, r,$ and $a$, yielding a grid of simulations summarized in Table \ref{tab:simulations}.
See Appendix \ref{sec:append-simulation} and Figure \ref{fig:sim-networks} for additional details.

\begin{table}[t!]
\processtable{} 
	{\begin{tabular}{|@{}m{12mm}|m{37mm}|m{19mm}|@{}} \hline \ Parameter & Meaning & Values \\ \hhline{|=|=|=|}
\ $\sizev$ & Number of variables & 40, 100, 200\\ \hline
\ $T$ & Time course length & 8 \\ \hline
\ $M$ & Number of time courses & 4 \\ \hline
\ $r$ & Fraction of original edges removed & 0.1, 0.5, 0.75, 1.0\\ \hline
\ $a$ & Fraction of spurious edges added &  0.1, 0.5, 0.75, 1.0 \\\hline
	\end{tabular}}{}
    \caption{These parameters define the grid of simulated datasets in our simulation study.
             There are $3 {\times} 4 {\times} 4 {=} 48$ distinct grid points.
						 For each one, we generate $K{=}5$ replicates
             for a total of 240 simulated datasets.
             The graph corruption parameters, $r$ and $a$, range from very little error (0.1) to total corruption (1.0).} 
        \label{tab:simulations}
\end{table}

\paragraph{Performance metrics.}
We are primarily interested in \Ourmethod{}'s ability to correctly recover the structure of the underlying signaling pathway.
The simulation study allows us to measure this in a setting where we have access to ground truth.
We treat this as a probabilistic binary classification task, where the method assigns an \emph{existence confidence} to each possible edge.
We measure classification performance using area under the precision-recall curve (AUCPR).
We use \emph{average precision} to estimate AUCPR, as opposed to the trapezoidal rule (which tends to be overly-optimistic, see \cite{davis-pr-roc-2006, flach-pr-2015}).

Our decision to use AUCPR is motivated by the sparseness of the graphs.
For sparse graphs the number of edges grows linearly with $\sizev$ while the number of possible edges grows quadratically. 
Hence, as $\sizev$ grows, the proportion of positive instances decreases and the classification task increasingly becomes a ``needle-in-haystack'' scenario.

Performance measurements on simulated data come with many caveats.
It's most instructive to think of simulated performance as a sanity check.
Since our data simulator closely follows our modeling assumptions, poor performance would suggest serious shortcomings in our method.

\subsection{HPN-DREAM network inference challenge evaluation}
\label{sec:dream}

We measure \Ourmethod's performance on experimental data by following the evaluation outlined by the HPN-DREAM Breast Cancer Network Inference Challenge \citep{2016-hill-community}.
Signaling pathways differ across contexts---e.g., cell type and environmental conditions.
The challenge is to infer these context-specific signaling pathways from time course data.

\paragraph{Dataset.}
The HPN-DREAM challenge provides phosphorylation time course data from 32 biological contexts.
These contexts arise from exposing 4 cell lines (BT20, BT549, MCF7, UACC812) to 8 stimuli.
For each context, there are approximately $M{=}4$ time courses, each about $T{=}7$ time points in length. 
Cell lines have differing numbers of phosphosite measurements (i.e., differing $\sizev$), ranging from 39 (MCF7) to 46 (BT20).

\paragraph{Prior knowledge.}
Participants in the original challenge were free to extract prior knowledge from any existing data sources.
As part of their analysis, the challenge organizers combined participants' prior graphs into a set of edge probabilities.
These \emph{aggregate priors} summarize the participants' collective knowledge.
They were not available to participants in the original challenge, but we use them in our analyses of HPN-DREAM data.
We provide them to each of the baseline methods (see Section \ref{sec:baselines}), so the resulting performance comparisons are fair.
We do not compare any of our scores to those listed by \cite{2016-hill-community} in the original challenge results.

\paragraph{Performance metrics.}
The HPN-DREAM challenge aims to score methods by their ability to capture causal relationships between pairs of variables.
It estimates this by comparing predicted \emph{descendant sets} against experimentally generated descendant sets.
More specifically, the challenge organizers exposed cells to AZD8055, an mTOR inhibitor, and observed the effects on other phosphosites. 
From this they determined a set of phosphosites \emph{downstream} of mTOR in the signaling pathway.
These include direct substrates of the mTOR kinase as well as indirect targets.

Comparing predicted descendants of mTOR against experimentally generated descendants of mTOR gives us a notion of \emph{false positives} and \emph{false negatives}.
As we vary a threshold on edge probabilities, the predicted mTOR descendants change, which allows us to make a receiver operating characteristic (ROC) curve.
We calculate the resulting area under the ROC curve (AUCROC) with the trapezoidal rule to score methods' performance on the HPN-DREAM challenge.
\cite{2016-hill-community} provide more details for this descendant set AUCROC scoring metric.
AUCROC is sensible for this setting since each descendant set contains a large fraction of the variables.
Sparsity is not an issue.

Because \Ourmethod{} is stochastic we run it $K{=}5$ times per context, yielding 5 AUCROC scores per context.
Meanwhile the baseline methods are all deterministic, requiring only one execution per context.
We use a simple terminology to compare \Ourmethod{}'s scores against those of other methods.
In a given context, we say \Ourmethod{} \emph{dominates} another method if its \emph{minimum} score exceeds that of the other method.
Conversely, we say the other method dominates \Ourmethod{} if its score exceeds \Ourmethod{}'s \emph{maximum} score.
This \emph{dominance} comparison has flaws---e.g., its results depend on the sample size $K$. 
However, it errs on the side of strictness and suffices as an aid for summarizing the HPN-DREAM evaluation results.

\subsection{Baseline pathway inference algorithms}
\label{sec:baselines}

Our evaluations compare \Ourmethod{} against a diverse set of baseline methods.

\paragraph{Exact DBN \ \citep{hill-bayesian-2012}.}
This method was an early inspiration for \Ourmethod{} and is most similar to \Ourmethod{}.
However, the exact DBN method encounters unique practical issues when we run it on real or simulated data.
The method's computational expense increases rapidly with problem size $\sizev$ and becomes intractable unless the ``max-indegree'' parameter is set to a small value.
For example, we found that the method used more than 32GB of RAM on problems of size $\sizev{=}100$, unless max-indegree was set ${\leq} 3$.
Furthermore, the exact DBN method only admits prior knowledge in the form of Boolean \emph{reference edges}, rather than continuous-valued edge confidences.
We overcame this by using a threshold to map edge confidences to 1 or 0.
We chose a threshold of 0.25 for the HPN-DREAM challenge evaluation because it yielded a reasonable number of prior edges.
We ran \citeauthor{hill-bayesian-2012}'s implementation using MATLAB 2018a.

\paragraph{\Funchisq{} \ \citep{zhang-funchisq-2013}.}
This method is based on the notion that two variables $X,Y$ have a causal relationship if there exists a \emph{functional dependence} $Y{=}f(X)$ between them.
It detects these dependencies using a chi-square test against the ``no functional dependence'' null hypothesis.
\Funchisq{} was a strong competitor in the HPN-DREAM challenge, despite the fact that it uses no prior knowledge.
In order to use \Funchisq{}, one must first discretize their time course data. 
We followed \citeauthor{zhang-funchisq-2013}'s recommendation to use 1D $k$-means clustering for discretization.
Detailed instructions are given in the HPN-DREAM challenge supplementary materials \citep{2016-hill-community}.
We used the \Funchisq{} (v2.4.9.1) and \texttt{Ckmeans.1d.dp} (v4.3.0) R packages.

\paragraph{LASSO.}
We included a variant of LASSO regression as a simple baseline.
It incorporates prior knowledge into the typical primal formulation:
\begin{equation*}
    \hat{\beta_j} = \argmin_{\beta} \left\{ \| X_{+,j} - B_j \beta \|_2^2 ~+~ \alpha \sum_{i=1}^V e^{-c_{ij}} |\beta_{i}| \right\}
\end{equation*}
where $c_{ij}$ is the prior confidence (either Boolean or real-valued) for edge $(i,j)$. 
That is, the method uses \emph{penalty factors} $e^{-c_{ij}}$ to discourage edges with low prior confidence.
The method selects LASSO parameters, $\alpha$, using the Bayesian Information Criterion described by \cite{zou-lasso-2007}.
We use \GLMNet{} \citep{friedman-glmnet-2010} via the 
\texttt{GLMNet.jl} Julia wrapper (v0.4.2).

\paragraph{Prior knowledge baseline.}
Our most straightforward baseline simply reports the prior edge probabilities, performing no inference at all.
Ideally, a Bayesian method should do no worse than the prior---new time course data should only \emph{improve} our knowledge of the true graph.
In reality, this improvement is subject to caveats about data quality and model fit.

\subsection{\Ourmethod{} software availability}
We provide the \Ourmethod{} code, distributed under a MIT license, via GitHub (\repolink{}) and archive it on Zenodo (\href{https://doi.org/10.5281/zenodo.3939287}{https://doi.org/10.5281/zenodo.3939287}).
It includes a Snakemake workflow \citep{koster-snakemake-2012} for our full evaluation pipeline, enabling the reader to reproduce our results. 
The code used in this manuscript corresponds to SSPS v0.1.1.

%% file: results.tex
\section{Results}
\label{sec:results}
We describe evaluation results from the simulation study and HPN-DREAM network inference challenge.
\Ourmethod{} competes well against the baselines, with superior scalability to other DBN-based approaches.

\subsection{Simulation study results}
\label{sec:sim-results}

We compare our method to the baselines listed in Section \ref{sec:baselines}.
We focus especially on the exact DBN method of \cite{hill-bayesian-2012}, as \Ourmethod{} shares many modeling assumptions with it.

\begin{table}
\processtable{} 
    {\begin{tabular}{|@{}m{12mm}|m{17mm}|m{22mm}|m{19mm}|@{}} \hline \ $\sizev$ & $\Nph$ & $\Neffph$ & MB per chain \\ \hhline{|=|=|=|=|}
\ 40 & 70000 & 400 & 500 \\ \hline
\ 100 & 9000 & 140 & 1200 \\ \hline
\ 200 & 3000 & 60 & 1000 \\ \hline
	\end{tabular}}{}
    \caption{Computational expense of \Ourmethod{} as a function of problem size $\sizev$.
             $N$ is the number of iterations completed by a Markov chain.
             $\Neff$ accounts for burnin and autocorrelation in the Markov chains, giving a more accurate sense of the method's progress.
             The last column gives the approximate memory footprint of each chain.
             The non-monotonic memory usage is an artifact of the chain termination conditions ($N{>}$100,000 or time ${>}12$ hours).
             }
    \label{tab:mcmc-expense}
\end{table}
        
\begin{table}
\processtable{} 
    {\begin{tabular}{|@{}m{8mm}|m{8mm}||m{15mm}|m{17mm}|@{}} \hline \ $\sizev$ & max indeg & ``linear'' & ``full'' \\ \hhline{|=|=|=|=|}
\ \multirow{4}{*}{40}  & 4 & 66s & 210s \\ 
        \cline{2-4}
\     & 5 & 770s & 3900s \\ 
        \cline{2-4}
\     & 6 & 6700s & \Timeout \\ 
        \cline{2-4}
\     & 7 & \OOM{} & \OOM{} \\ \hline
\ \multirow{2}{*}{100} & 3 & 250s & 520s \\ 
        \cline{2-4}
\     & 4 & \OOM{} & \OOM{} \\ \hline
\ \multirow{2}{*}{200} & 2 & 53s & 140s \\ 
        \cline{2-4}
\     & 3 & \OOM{} & \OOM{} \\ \hline
	\end{tabular}}{}
    \caption{Computational expense of the exact DBN method of \cite{hill-bayesian-2012} measured in CPU-seconds, as a function of problem size $\sizev$ and various parameter settings.
             The method imposes an in-degree constraint on each vertex, shown in the ``max indeg'' column.
             The columns ``linear'' and ``full'' correspond to different \emph{regression modes}, i.e., which interaction terms are included in the DBN's conditional probability distributions.
             ``\OOM{}'' (Out Of Memory) indicates that the method exceeded a 32GB memory limit.
             ``\Timeout{}'' indicates that the method failed to complete within 12 hours.
             }
    \label{tab:hill-expense}
\end{table}

\paragraph{Computational expense.}
Because \Ourmethod{} uses MCMC, the user may allow it to run for an arbitrary amount of time.
With this in mind, we summarize \Ourmethod{}'s timing with two numbers:
\primo{} $\Nph$, the number of MCMC samples per CPU-hour; and
\secundo{} $\Neffph$, the \emph{effective} number of samples per CPU-hour.
We also measure the memory footprint per Markov chain, subject to our termination conditions.
We measured these numbers for each simulation in our grid (see Table \ref{tab:simulations}).

Table \ref{tab:mcmc-expense} reports average values of $\Nph$, $\Neffph$, and memory footprint for each problem size.
As we expect, $\Nph$ and $\Neffph$ both decrease approximately with the inverse of $\sizev$.
In contrast, the non-monotonic memory usage requires more explanation.
It results from two causes: \primo{} our termination condition and
\secundo{} the sparse data structures we use to store samples.
On small problems ($\sizev{=}40$), the Markov chain terminates at a length of 100,000---well within the 12-hour limit.
On larger problems ($\sizev{=}100$ or $200$) the Markov chain terminates at the 12-hour timeout.
This accounts for the 500MB gap between small and large problems.
The \emph{decrease} in memory usage between $\sizev{=}100$ and $200$ results from our sparse representations for samples.
Roughly speaking, the sparse format only stores \emph{changes} in the variables.
So the memory consumption of a Markov chain depends not only on $\sizev$, but also on the \emph{acceptance rate} of the Metropolis-Hastings proposals.
The acceptance rate is smaller for $\sizev{=}200$, yielding a net decrease in memory usage.

Recall that \Ourmethod{} differs from more traditional MCMC approaches by nature of its parent set proposal distribution, which is specially designed for sparse graphs (see Section \ref{sec:inference}).
When we modify \Ourmethod{} to instead use a na\"ive uniform graph proposal, we see a striking difference in sampling efficiency.
The uniform graph proposal distribution attains $\Neffph$ of 100, 10, and 0.2 for  $\sizev{=}40,100,$ and $200$, respectively---drastically smaller than those listed in Table \ref{tab:mcmc-expense} for the parent set proposal. 
It's possible that the traditional proposal could achieve higher $\Neffph$ by simply running faster.
However, the more important consideration is how $\Neffph$ changes with $\sizev$.
Our parent set proposal distribution's $\Neffph$ decays approximately like $O(1/\sizev)$.
This is better than what we might expect from a simple analysis (Appendix \ref{sec:append-proposal}).
Meanwhile, the traditional proposal distribution's $\Neffph$ decays faster than $O(1/\sizev^4)$.
This gap between $O(1/|V|)$ and $O(1/|V|^4)$ sampling efficiencies makes an enormous difference on large problems.

Table \ref{tab:hill-expense} summarizes the computational expense of the exact DBN method \citep{hill-bayesian-2012}.
The method quickly becomes impractical as the problem size grows, unless we enforce increasingly strict in-degree restrictions.
In particular, the exact DBN method's memory cost grows exponentially with its ``max in-degree'' parameter.
The growth becomes increasingly sharp with problem size.
When $\sizev{=}200$, increasing the maximum in-degree from 2 to 3 makes the difference between terminating in ${<}1$ minute and exceeding 32GB of memory.
Such low bounds on in-degree are unrealistic, and will likely result in poor inference quality. 
In comparison, \Ourmethod{} makes no constraints on in-degree, and its memory usage scales well with problem size.

The other baseline methods---\Funchisq{} and LASSO---are much less computationally expensive.
Both finish in seconds and require less than 100MB of memory for each simulated task.
This highlights the computationally intense nature of Bayesian approaches.
Not every scenario calls for Bayesian inference.
However, Bayesian inference is valuable in scientific settings where we're concerned with uncertainty quantification.

    \begin{figure}[t!]
        \includegraphics[scale=0.5]{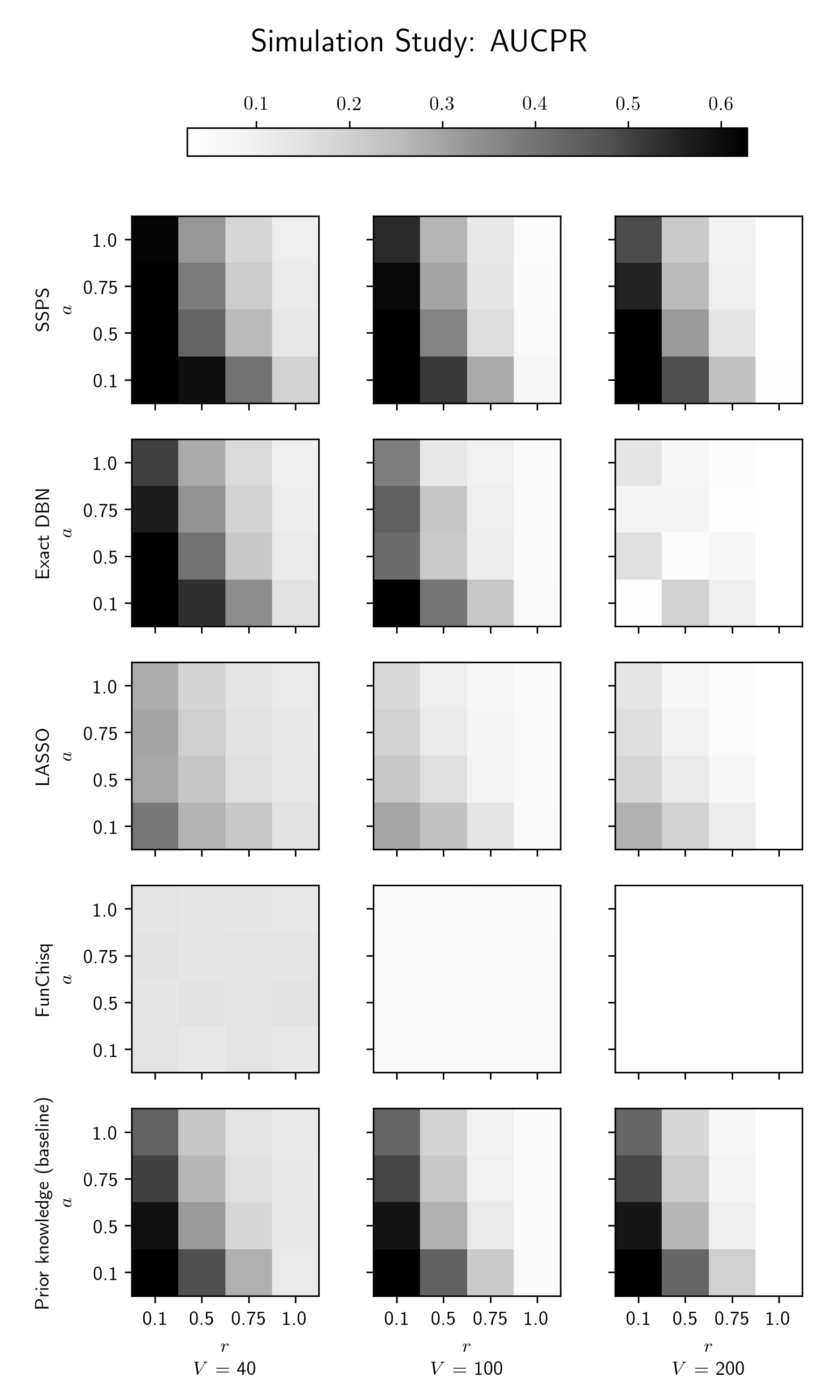}
        \caption{Heatmap of AUCPR values from the simulation study.
                 Both DBN-based techniques (\Ourmethod{} and the exact method) score well on this, since the data is generated by a DBN.
                 On large problems the exact DBN method needs strict in-degree constraints, leading to poor prediction quality.
                 LASSO and \Funchisq{} both perform relatively weakly.
								 See Figure \ref{fig:curves-sim} for representative ROC and PR curves.
                }
        \label{fig:aucpr-heat}
    \end{figure}

\paragraph{Predictive performance.}    
The simulation study provides a setting where we have access to ``ground truth''---the true simulated graph.
We use AUCPR to score each method's ability to recover the true graph's edges.

Figure \ref{fig:aucpr-heat} shows the AUCPR scores for our grid of simulations.
Each heat map shows AUCPR as a function of graph corruption parameters, $r$ and $a$.
The heat maps are arranged by method and problem size $\sizev$.
Each AUCPR value is an average over 5 replicates.
\Ourmethod{} maintains fairly consistent performance across problem sizes.
In contrast, the other methods' scores decrease with problem size.
For the exact DBN method, this is partially due to the small in-degree constraints imposed on the large problems.
It is forced to trade model accuracy for tractability.

\begin{figure}[t]
        \includegraphics[scale=0.5]{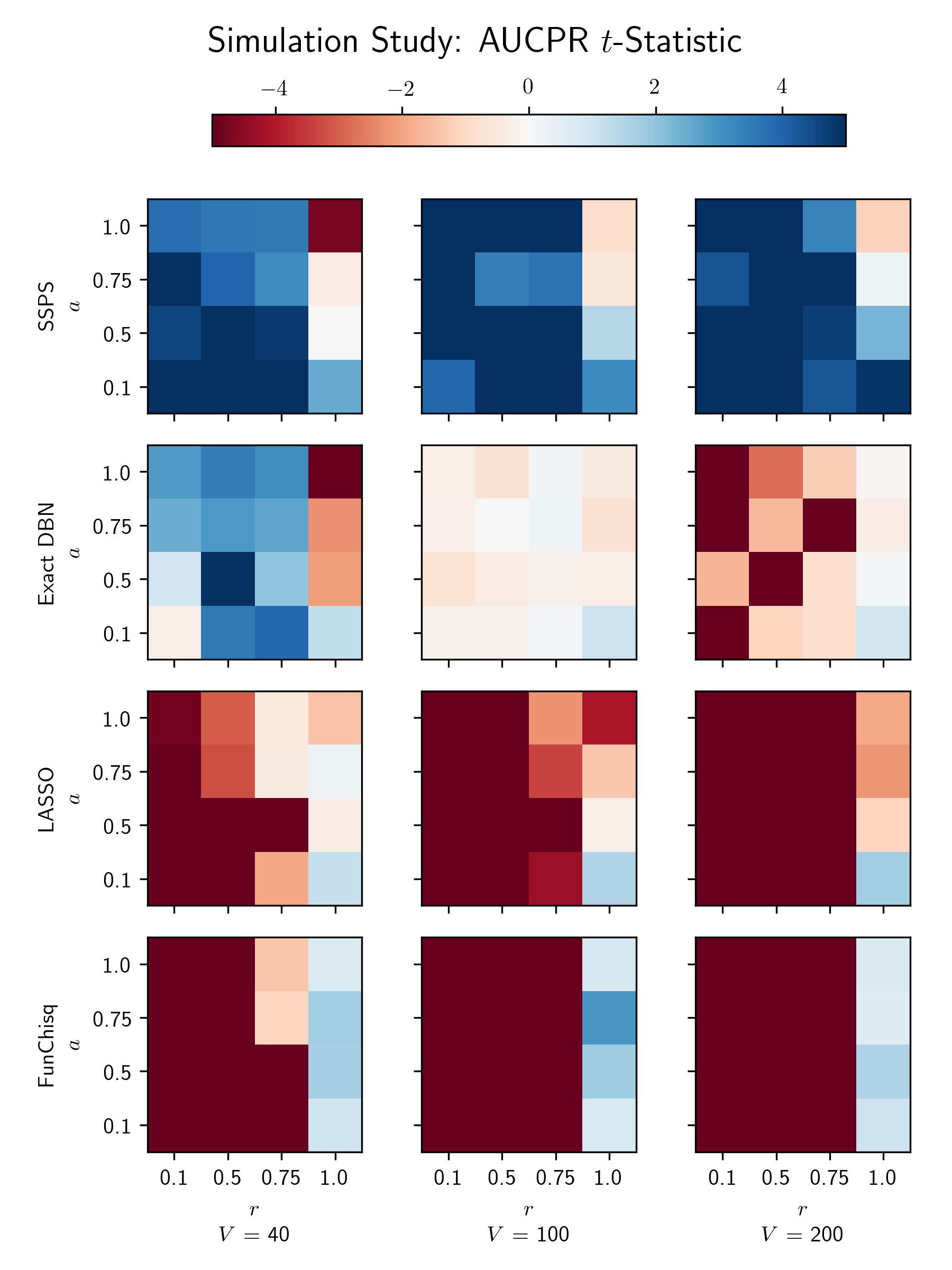}
        \caption{Heatmap of differential performance against the prior knowledge, measured by AUCPR paired $t$-statistics.
                 \Ourmethod{} consistently outperforms the prior knowledge across problem sizes and shows robustness to errors in the prior knowledge.
                 }
        \label{fig:t-heat}
    \end{figure}

Figure \ref{fig:t-heat} reveals further insights into these results.
It plots \emph{differential} performance with respect to the prior knowledge, in a layout analogous to Figure \ref{fig:aucpr-heat}.
Specifically, it plots the $t$-statistic of each method's AUCPR, paired with the prior baseline's AUCPR.
Whenever the prior graph has some informative edges, \Ourmethod{} outperforms the prior.
On the other hand, \Ourmethod{}'s performance deteriorates whenever the prior contains \emph{no} true edges (i.e., $r{=}1$).
Under those circumstances \Funchisq{} may be a better choice. 
Since it doesn't rely on prior knowledge at all, it outperforms the other methods when the prior is totally corrupted.
However, we expect that in most realistic settings there exists partially-accurate prior knowledge, in which case we expect \Funchisq{} to perform worse than \Ourmethod{}.

These results confirm \Ourmethod{}'s ability to identify the true network, given partially-accurate prior knowledge and time series data consistent with the modeling assumptions.
\Ourmethod{} is fairly robust with respect to the prior's quality and has consistent performance across different problem sizes.

\subsection{HPN-DREAM challenge results}
\label{sec:dream-results}

\begin{figure*}[!htbp]
        \centering
        \includegraphics[scale=0.5]{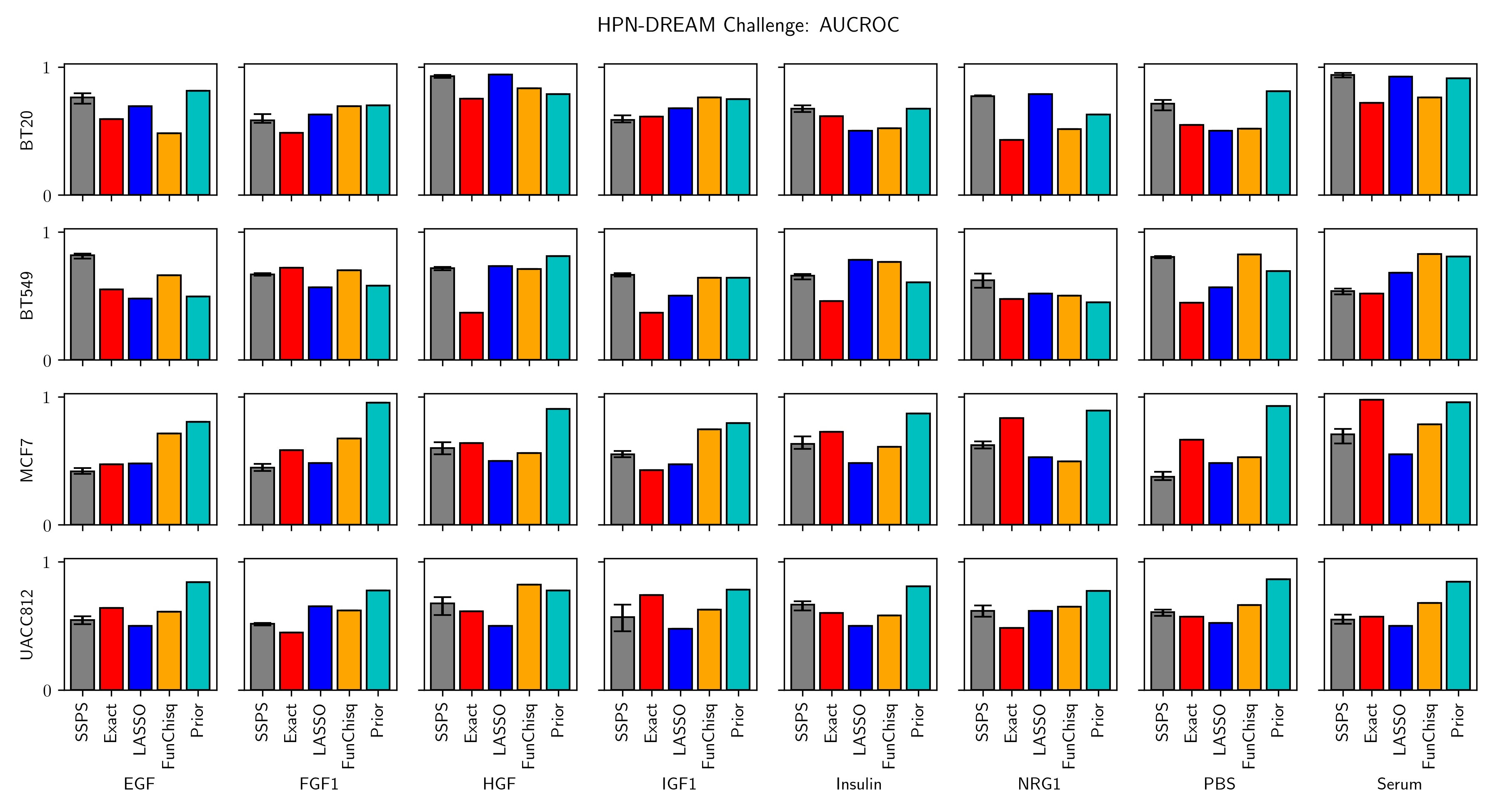}
        \caption{Methods' performances across contexts in the HPN-DREAM Challenge.
                 MCMC is stochastic, so we run \Ourmethod{} 5 times; the error bars show the range of AUCROC scores. 
                 The other methods are all deterministic and require no error bars.
								 See Figure \ref{fig:dream-networks} for example predicted networks, Figure \ref{fig:aucpr-bar} for AUCPR scores, and Figure \ref{fig:curves-dream} for representative ROC and PR curves.
                 }
        \label{fig:dream-bar}
    \end{figure*}

We evaluated \Ourmethod{} on experimental data from the HPN-DREAM challenge.
The challenge includes time series phosphorylation data from 32 biological contexts: 8 stimuli applied to 4 breast cancer cell lines.
Methods are scored on their ability to correctly identify the experimentally derived descendants of mTOR.
Figure \ref{fig:dream-bar} shows bar charts comparing the methods' AUCROC scores in each context.
Appendix \ref{sec:append-dream} provides additional details.

\Ourmethod{} performs satisfactorily on this task overall.
Employing terminology from Section \ref{sec:dream}, \Ourmethod{} dominates the exact DBN method in 18 of the 32 contexts, whereas the exact DBN method dominates \Ourmethod{} in only 9 contexts.
Meanwhile, \Ourmethod{} dominates \Funchisq{} in 11 contexts and is dominated by \Funchisq{} in 15.
This is not surprising because \Funchisq{} was among the top competitors in the original challenge. 
LASSO, on the other hand, performs poorly.
\Ourmethod{} dominates LASSO in 18 contexts and is dominated in only 6.

More puzzling is the strong performance of the prior knowledge baseline.
\Ourmethod{} dominates the aggregate prior in only 9 contexts and is dominated in 21.
This is not isolated to our method.
\Funchisq{} outperforms and is outperformed by the prior knowledge in 11 and 21 contexts, respectively.
The aggregate prior's strong performance is consistent with the results from the original HPN-DREAM challenge;
this prior outperformed all individual challenge submissions \citep{2016-hill-community}.
Even though the aggregate prior gives identical predictions for each context and totally ignores the time course data,
it still attains better performance than the other methods.
This suggests either \primo{} the data is relatively uninformative or
\secundo{} the evaluation metric based on mTOR's descendants isn't sufficiently precise to measure context-specific performance.
We suspect the latter, because \Funchisq{} uses no prior knowledge but was the top performer
in the HPN-DREAM challenge's \emph{in silico} tasks.
An evaluation based on one node's descendants is not as discriminative as an evaluation of the directed edges.
Many different directed graphs can have equivalent or similar mTOR descendants.
However, it is experimentally impractical to generate the context-specific gold standard networks that would be required for a more precise edge-based evaluation.

%% file: discussion.tex
\section{Discussion}

We presented \Ourmethod{}, a signaling pathway reconstruction technique based on DBN structure estimation.
It uses MCMC to estimate the posterior probabilities of directed edges, employing a parent set proposal distribution specially designed for sparse graphs.
\Ourmethod{} is a Bayesian approach.
It takes advantage of prior knowledge with edge-specific confidence scores and can provide uncertainty estimates on the predicted pathway relationships, which are valuable for prioritizing experimental validation.

\Ourmethod{} scales to large problems more efficiently than past DBN-based techniques.
On simulated data, \Ourmethod{} yields superior edge predictions with robustness to flaws in the prior knowledge.
Our HPN-DREAM evaluation shows \Ourmethod{} performs comparably to established techniques on a community standard task.
It is difficult to make stronger statements in the HPN-DREAM setting because the prior knowledge baseline performs so well and we can only evaluate the predicted mTOR descendants, not the entire pathway.
However, \Ourmethod{}'s scalability among Bayesian methods, strong results in the simulation, and competitive performance in the HPN-DREAM challenge make it an attractive option for further investigation of real phosphorylation datasets.

There are several potential limitations of \Ourmethod{} relative to alternative pathway signaling models.
Prior knowledge is not available in some organisms or biological conditions, reducing one advantage of our Bayesian approach.
Although \Ourmethod{} is more scalable than related DBN techniques, it would struggle to scale to proteome-wide phosphoproteomic data measuring thousands of phosphosites.
For large datasets, we recommend running \Ourmethod{} on a pruned version that includes only the highest intensity or most variable phosphosites.
\Ourmethod{}, like most DBN techniques, models only observed variables.
It will erroneously exclude important pathway members, such as scaffold proteins, that are not phosphorylated.
Latent variable models or background network-based algorithms are better suited for including unphosphorylated proteins in the pathway.
Background network methods can also impose global constraints on the predicted pathway structure, such as controlling the number of connected components or proteins' reachability from relevant receptors \citep{2018-koksal-tps}.

There are many possible ways to improve \Ourmethod.
For example, it could be extended to jointly model related pathways in a hierarchical fashion, similar to \cite{2014-oates-joint} and \cite{2017-hill-context}. 
Alternatively, \Ourmethod{} could be modified to accommodate causal assumptions via Pearl's intervention operators; 
see the model of \cite{2015-spencer-interventional} for a relevant example.
Combining temporal and interventional data \citep{cardner-inferring-2019} is another rich area for future work.
On the algorithmic side, we could improve our MCMC procedure by adaptively tuning the parameters of its proposal distributions, as described by \cite{gelman-text-2014}.  
Because \Ourmethod{} is a probabilistic program, it is naturally extensible.

%% file: appendices.tex
\section*{Appendix}

\subsection{Model formulation details}
\label{sec:append-model}
We provide additional information about our graph prior and marginal likelihood function.
We also describe some implications of \Ourmethod{}'s model assumptions.

\paragraph{Derivation of graph prior (Equation \ref{eq:new-prior}).} We step through a more detailed derivation of \Ourmethod{}'s new graph prior.
We begin with the original graph prior (Equation \ref{eq:oldprior}) and rewrite it in terms of the edge existence variables $Z$:
\begin{align}
    P(G|G^\prime, \lambda) & \propto \exp\left( -\lambda | E(G) \setminus E(G^\prime)| \right) \nonumber \\[4pt] 
                       & = \exp\left(-\lambda \sum_{(i,j) \notin E(G^\prime)} z_{ij} \right) \label{eq:secondline} \\[4pt]
                       & = \prod_{(i,j) \notin E(G^\prime)}\!\!\! e^{-\lambda z_{ij} } \nonumber \\
                       & = \prod_{(i,j) \notin E(G^\prime)}\!\!\! \left(e^{-\lambda}\right)^{z_{ij}} \nonumber \\[4pt]
                       & \propto \left( \frac{1}{1 + e^{-\lambda}} \right)^{V^2 - |E(G^\prime)|} \cdot \prod_{(i,j) \notin E(G^\prime)}\!\!\! \left(e^{-\lambda}\right)^{z_{ij}} \nonumber \\[4pt]
                       & = \prod_{(i,j) \notin E(G^\prime)}\!\!\! \left( \frac{1}{1 + e^{-\lambda}} \right) \left(e^{-\lambda}\right)^{z_{ij}} \nonumber \\[4pt]
                       & = \prod_{(i,j) \notin E(G^\prime)}\!\!\! \left( \frac{1}{1 + e^{-\lambda}} \right)^{1 - z_{ij}} \left( \frac{e^{-\lambda}}{1 + e^{-\lambda}} \right)^{z_{ij}} \label{eq:lastline}
\end{align}
Equation \ref{eq:lastline} shows the original prior is in fact a product of independent Bernoulli variables---the edge existence variables $z_{ij}$.
Equation \ref{eq:lastline} explicitly assigns probability to the edges \emph{not} contained in $E(G^\prime)$.
However, it also implicitly assigns uniform probability to every edge \emph{contained} in $E(G^\prime)$.
We deduce that they are $\Bernoulli(0.5)$ variables, allowing us to write the prior $P(Z~|~G^\prime, \lambda)$ in the following form:
\begin{equation}
\prod_{(i,j) \in E(G^\prime)}\!\!\! \left(\frac{1}{2}\right)^{z_{ij}}\!\left(\frac{1}{2}\right)^{1 {-} z_{ij}}\!\!\!\!\! \prod_{(i,j) \notin E(G^\prime)} \left( \frac{e^{-\lambda}}{1 {+} e^{-\lambda}} \right)^{z_{ij}}\! \left( \frac{1}{1 {+} e^{-\lambda}} \right)^{1 {-} z_{ij}}
    \label{eq:bern-prior}
\end{equation}
just as shown in the main text.

Now we modify the prior to use continuous-valued edge confidences $c_{ij}$ instead of Boolean reference edges $E(G^\prime)$.
Intuitively, we want to restate Equation \ref{eq:bern-prior} as a single product over all $Z$ variables, rather than two separate products.
Our goal is to find a function $q(c_{ij})$ such that
$$ P(Z~|~C, \lambda) = \prod_{(i,j)} q(c_{ij})^{z_{ij}} (1 - q(c_{ij}))^{1 - z_{ij}}.$$
However, in order to remain consistent with the original prior $q(c_{ij})$ ought to be monotone-increasing and satisfy these criteria:
$$q(0) = e^{-\lambda} / (1 + e^{-\lambda}) \hspace{5em} \text{and} \hspace{5em} q(1) = 1/2.$$
It turns out that choosing
$$q(c_{ij}) = \frac{e^{-\lambda}}{e^{-c_{ij} \lambda} + e^{-\lambda}}$$
satisfies these requirements.
This brings us to Equation \ref{eq:prior-conts} of the main text.

From there, it is straightforward to replace the single shared $\lambda$ variable with a set of vertex-specific $\Lambda$ variables and arrive at Equation \ref{eq:new-prior}.

\paragraph{Marginal likelihood function details.} 
Equation \ref{eq:marglik} is obtained by \primo{} using a Gaussian DBN as the likelihood function for $G$, \secundo{} assuming certain prior distributions for the DBN parameters, and \tertio{} integrating the DBN parameters out.
Specifically, let $\beta_j$ and $\sigma_j^2 ~ ~ \forall j \in \rangev$ be the DBN's weight and noise parameters, respectively.
We assume an improper prior $\sigma_j^2 \propto 1 / \sigma_j^2$ for the noise and a Gaussian prior for the weights:
$$ \beta_j | \sigma_j^2 \sim \mathcal{N}\left(0, T\sigma_j^2(B_j^\trans B_j)^{-1}\right).$$
In other words, SSPS uses an improper joint prior $P(\beta_j, \sigma_j^2) = P(\beta_j | \sigma_j^2) P(\sigma_j^2)$ with $P(\sigma_j^2) {\propto} 1/\sigma_j^2$. 
This choice allows $\beta_j$ and $\sigma_j^2$ to be marginalized, yielding Equation \ref{eq:marglik}.

The power $-|\parents_G(j)|/2$ in Equation \ref{eq:marglik} is correct when the DBN only uses linear terms.
Recall that $B_j$ may in general contain columns of nonlinear interactions between parent variables.
When that is true, the quantity $|\parents_G(j)|$ should be replaced by the \emph{width of} $B_j$.
We elide this detail in the main text for brevity.
Our implementation uses the correct exponent.

Our implementation of the marginal likelihood function employs least recently used caching to reduce redundant computation.
Code profiling shows that this yields a substantial improvement to efficiency.
For additional in-depth discussion of Equation \ref{eq:marglik}, we recommend the supplementary materials of \cite{hill-bayesian-2012}.

\paragraph{Additional insights about \Ourmethod{}'s model assumptions.}
\Ourmethod{}'s model has interesting properties that could lead to method improvements.
For example, when we replace the shared $\lambda$ variable with vertex-specific $\Lambda$ variables, the model effectively becomes a set of $|V|$ independent models.
The plate notation in Figure \ref{fig:new-model} makes this clear; 
$X_-$ is the only shared variable, and it's fully observed.
This has algorithmic implications.
For example, future versions of \Ourmethod{} could parallelize inference at the vertex level, allocating more resources to the parent sets that converge slowly.

In the course of deriving Equation \ref{eq:lastline}, we showed that our prior is a log-linear model over edge features.
Equation \ref{eq:secondline} shows this most clearly.
Future versions of \Ourmethod{} could use the expressiveness of log-linear densities over higher-order graph features to capture richer forms of prior knowledge.

\subsection{Parent set proposal details}
\label{sec:append-proposal}
A key component of \Ourmethod{} is its novel \emph{parent set proposal distribution}.
We motivate its design and discuss its computational complexity in greater detail. 

\paragraph{Parent sets instead of edges.}
The marginal likelihood (Equation \ref{eq:marglik}) is a function of the graph $G$.
However, it depends on $G$ only via its \emph{parent sets}, which are encoded in the matrices $B_j$.
Accordingly, \Ourmethod{} represents $G$ by storing a list of parents for each vertex.

It makes sense to use a proposal distribution that operates directly on \Ourmethod{}'s internal parent set representation.
This motivates our choice of the \addparent{}, \removeparent{}, and \swapparent{} proposals listed in Section \ref{sec:inference}.
There is a natural correspondence between 
\primo{} likelihood function, 
\secundo{} data structure, and
\tertio{} proposal distribution.

\paragraph{Sampling efficiency.}
We provide some intuition for the parent set proposal's superior sampling efficiency.
Let $z_{ij}$ be a particular edge existence variable.
The estimate for $z_{ij}$ converges quickly if MCMC updates $z_{ij}$ frequently.
Hence, as a proxy for sampling efficiency, consider the number of times $z_{ij}$ gets updated per unit time.
We decompose this quantity into three factors:
$$
\frac{z_{ij} \text{ updates}}{\text{unit time}} = \epsilon  \cdot \tau \cdot \alpha 
$$
where
$$
\epsilon = \frac{\text{graph proposals}}{\text{unit time}}  \hspace{4em}
\tau = \frac{z_{ij} \text{ proposals}}{\text{graph proposal}} 
$$
\vspace{-3pt}
$$
\alpha = z_{ij} \text{ acceptance probability}
$$
In other words, $\epsilon$ is the time efficiency of the proposal distribution.
The factor $\tau$ is the probability that a given proposal \emph{touches} $z_{ij}$. 
Lastly, $\alpha$ is the proposal's Metropolis-Hastings acceptance probability.

For a given proposal distribution, we're interested in how these factors depend on $|V|$.
For simplicity of analysis, assume the Markov chain is in a typical state where the graph is sparse: $|E(G)| = O(|V|)$.

For the parent set proposal, execution time has no dependence on $|V|$ and hence $\epsilon~{=}~O(1)$.
Recall that the parent set proposal resides in an outer loop, which iterates through all $|V|$ vertices.
It follows that for any particular proposal there is a $1/|V|$ chance that it acts on vertex $j$.
After choosing vertex $j$, there is on average a $O(1/|V|)$ chance that the proposal affects $z_{ij}$.
This follows from the sparsity of the graph: vertex $i$ is typically a non-parent of $j$ and the probability of choosing it via an \addparent{} or \swapparent{} action is $O(1/|V|)$.
Hence, the parent set proposal has a probability $\tau {=} O(1/|V|^2)$ of choosing $z_{ij}$.
Lastly, the acceptance probability $\alpha$ has no dependence on $|V|$ and therefore $\alpha = O(1)$.
The product of these factors gives an overall sampling efficiency of $O(1/|V|^2)$ for the parent set proposal.

For the uniform graph proposal, $\epsilon$'s complexity depends on the particular implementation.
For sake of generosity we assume an efficient implementation with $\epsilon~{=}~O(1)$.
The proposal chooses uniformly from $O(|V|^2)$ actions: add-, remove-, or reverse-edge. 
The probability of choosing one that affects $z_{ij}$ is $\tau = O(1/|V|^2)$.
Recall that the marginal likelihood decreases steeply with parent set size.
It follows that add-edge actions will typically have low acceptance probability.
Since the graph is sparse, add-edge actions are overwhelmingly probable; the probability of \emph{not} landing on one is $O(1/|V|^2)$.
If we assume the acceptance probability is high for remove-edge and reverse-edge actions, (i.e., they are accepted whenever they're proposed), then this suggests $\alpha = O(1/|V|^2)$, averaged over many proposals.
The product of these factors suggests a sampling efficiency that decays like $O(1/|V|^4)$.

This gap between $O(1/|V|^2)$ and $O(1/|V|^4)$ sampling efficiencies explains most of the difference that we saw in Section \ref{sec:sim-results}.
A more detailed analysis may reveal why the parent set proposal attains sampling efficiencies closer to $O(1/|V|)$ in practice.

\subsection{Simulation study details}
\label{sec:append-simulation}

\begin{figure*}
    \includegraphics[scale=1.0]{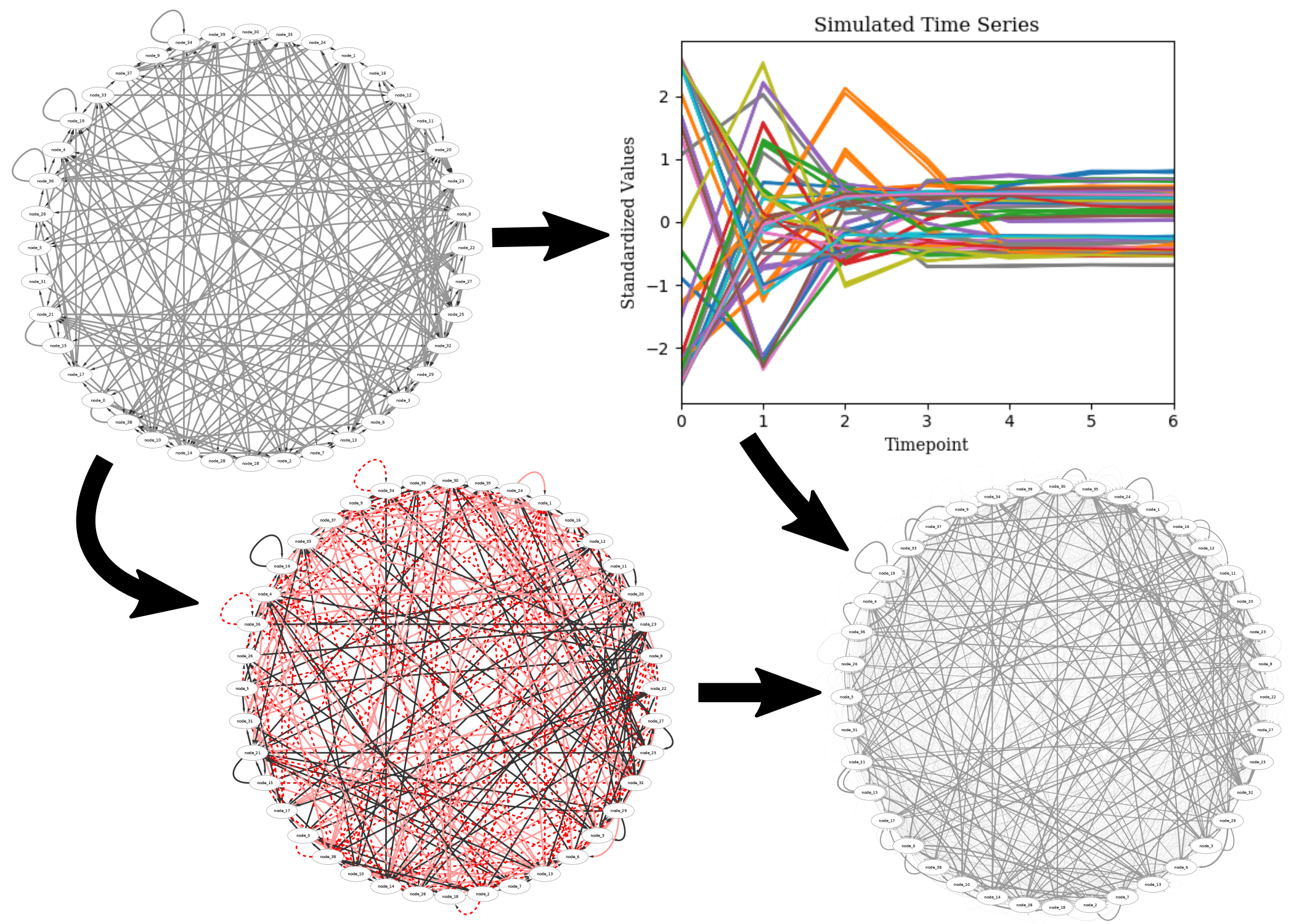}
    \caption{A schematic of the simulation study.
             We randomly generate a true network (upper left) and  
             use it to simulate a time series dataset (upper right).
             We corrupt the true network by adding and removing edges (lower left);
             solid red edges have been added, dashed red edges have been removed, and black edges are original.
             This corrupted network serves as partially inaccurate prior knowledge for the inference techniques.
             Each technique produces a predicted network (lower right) by assigning a score to each possible edge.
						 The predicted network is evaluated with respect to the true network.
             }
    \label{fig:sim-networks}
\end{figure*}

\begin{figure*}
    \centering
    \textbf{ROC Curves}\\
    \includegraphics[scale=0.42]{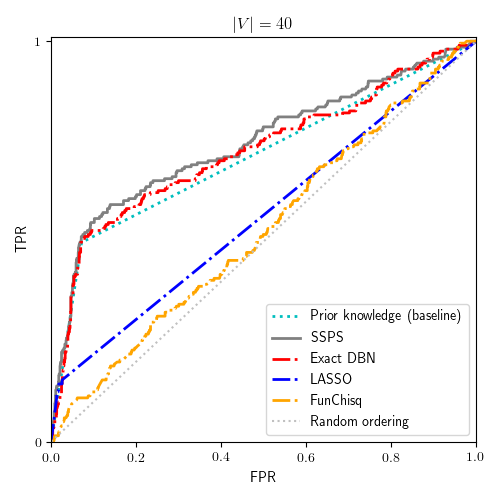}
    \includegraphics[scale=0.42]{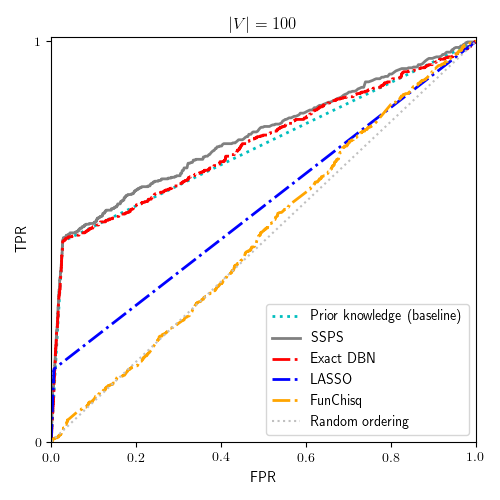}
    \includegraphics[scale=0.42]{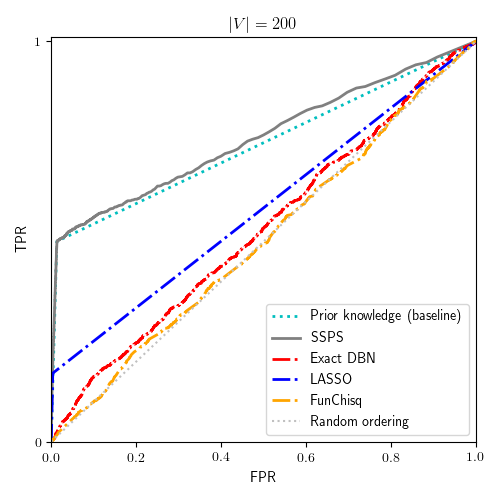}\\
    \vspace{4pt}
    \textbf{PR Curves}\\
    \includegraphics[scale=0.33]{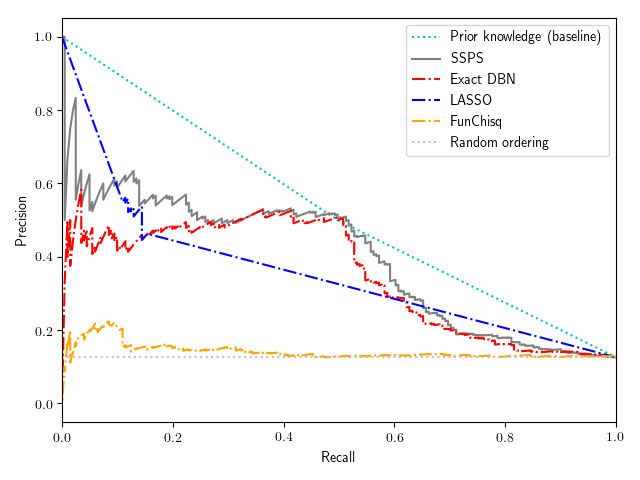}
    \includegraphics[scale=0.33]{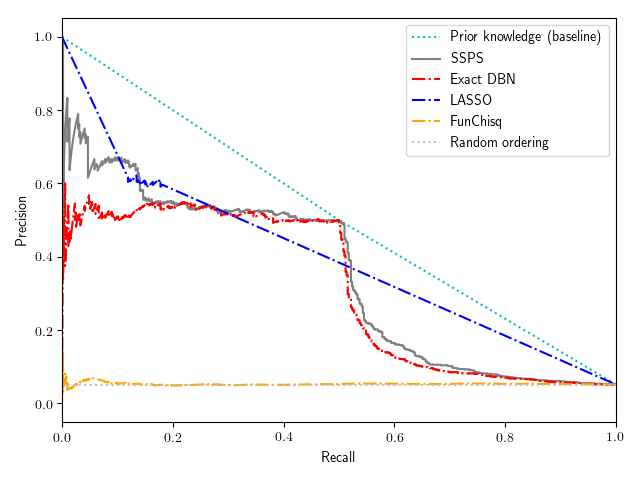}
    \includegraphics[scale=0.33]{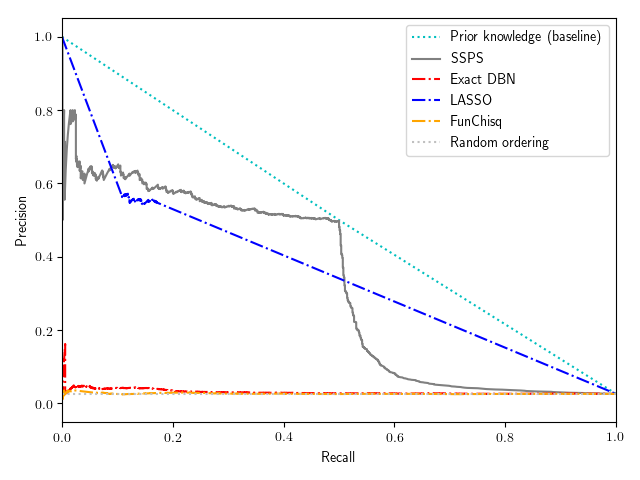}
    \caption{Representative ROC curves (top) and PR curves (bottom) from the simulation study.
             We show curves for three different simulations: $|V|=40,100,$ and $200$ (left, middle, right respectively).
             Each of these simulations used corruption parameters $r=a=0.5$.}
    \label{fig:curves-sim}
\end{figure*}

We give additional details about the simulation study's methodology and results.

\paragraph{Simulation process.}
The simulation process described in Section \ref{sec:simulation} differs from \Ourmethod{}'s modeling assumptions in several ways.
Recall that the simulator constructs a DBN to generate time series data.
This simulated DBN employs nonlinear interaction terms.
The simulator assumes that the data at each timepoint is a \emph{cubic} function of the data at the previous timestep.
In contrast, all of our analyses ran \Ourmethod{} with an assumption of \emph{linear} dependencies.
In other words, the data contained complexities that \Ourmethod{} was unable to capture.
\Ourmethod{}'s performance in the simulation study suggests that it has some robustness to modeling assumption mismatches.

We provide an illustration of the simulation process in Figure \ref{fig:sim-networks}.
It is interesting to notice that the simulated networks do not resemble directed acyclic graphs (DAGs) in any way.
They do not have any sense of directionality.
Contrast this with the biological graphs shown in Figure \ref{fig:dream-networks}.
Strictly speaking these are not DAGs, but they do have an overall direction.
Some vertices are source-like, and others are sink-like.
Future simulations and models could be more biologically realistic if they incorporated this kind of structure. 

\paragraph{Simulation study results.}
Figure \ref{fig:curves-sim} gives some representative ROC and PR curves from the simulation study.
On problem sizes up to $|V|=100$, \Ourmethod{} and the exact DBN method yield similar curves in both ROC and PR space---though \Ourmethod{}'s curves clearly dominate.
On larger problems the exact DBN method's performance quickly deteriorates. 
Computational tractability requires the exact method to impose highly restrictive in-degree constraints.
These observations are consistent with the heatmaps of Figures \ref{fig:aucpr-heat} and \ref{fig:t-heat} in the main text.

\subsection{HPN-DREAM challenge details}
\label{sec:append-dream}

We provide additional details for the methodology and results of the HPN-DREAM challenge evaluation.

\paragraph{Data preprocessing.}
The HPN-DREAM challenge data needed to be preprocessed before it could be used by the inference methods.
The choices we made during preprocessing most likely affected the inference results.

Many of the time series contain duplicate measurements.
We managed this by simply averaging the duplicates.
We log-transformed the time series since they were strictly positive and some methods (\Ourmethod{} and exact DBN) assume normality.
This probably made little difference for \Funchisq{}, which discretizes the data as part of its own preprocessing.

\begin{figure*}
    \includegraphics[scale=0.85]{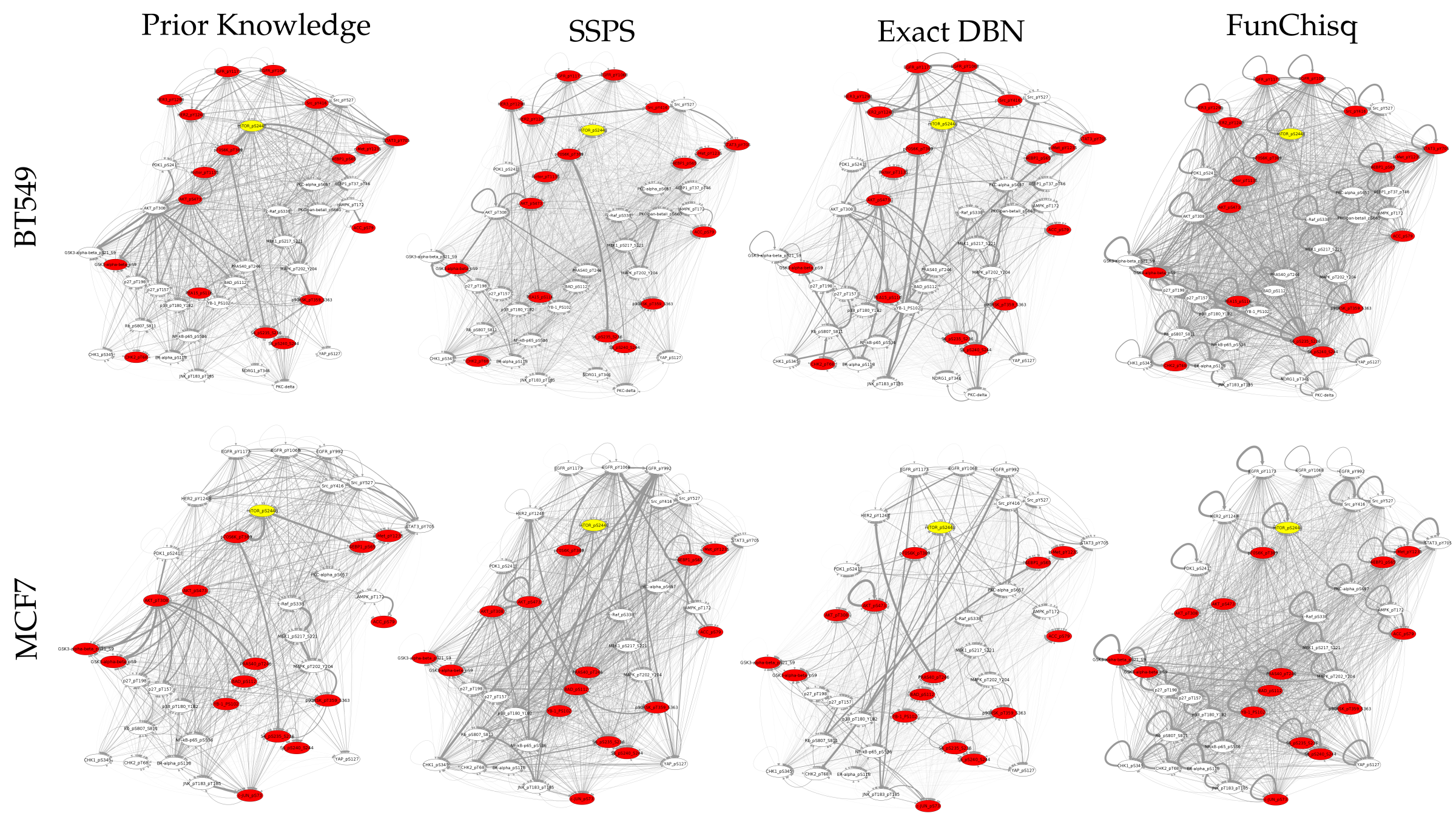}
    \caption{Prior and predicted pathways from the HPN-DREAM challenge.
             We show pathways from two contexts: cell lines BT549 (top row) and MCF7 (bottom row).
             The stimulus is EGF for both contexts.
             SSPS attained the best AUCROC of all methods in the (BT549, EGF) context
             and the worst in the (MCF7, EGF) context.
             The yellow node is mTOR; red nodes are the experimentally generated (``ground truth'') descendants of mTOR.
             }
    \label{fig:dream-networks}
\end{figure*}

\paragraph{Predicted networks.} 
Figure \ref{fig:dream-networks} visualizes networks from two biological contexts in the HPN-DREAM challenge evaluation.
This gives a sense of how the different inference methods' predictions differ from each other.
All of the predicted networks are fairly different, though the SSPS and exact DBN predictions are more similar to each other than they are to \Funchisq{}.
\Funchisq{} predicts more self-edges than the other methods.

In the BT549 cell line, the experimentally detected mTOR descendants include receptor proteins that would traditionally be considered upstream of mTOR in the pathway.
The experimental results are reasonable due to the influence of feedback loops in signaling pathways.
However, the number and positioning of the mTOR descendants highlights the differences between the coarse HPN-DREAM challenge evaluation, which is based on reachability in a directed graph, and the more precise evaluation in our simulation study, where we have the edges in the ground truth network.

\begin{figure*}
    \centering
    \includegraphics[scale=0.5]{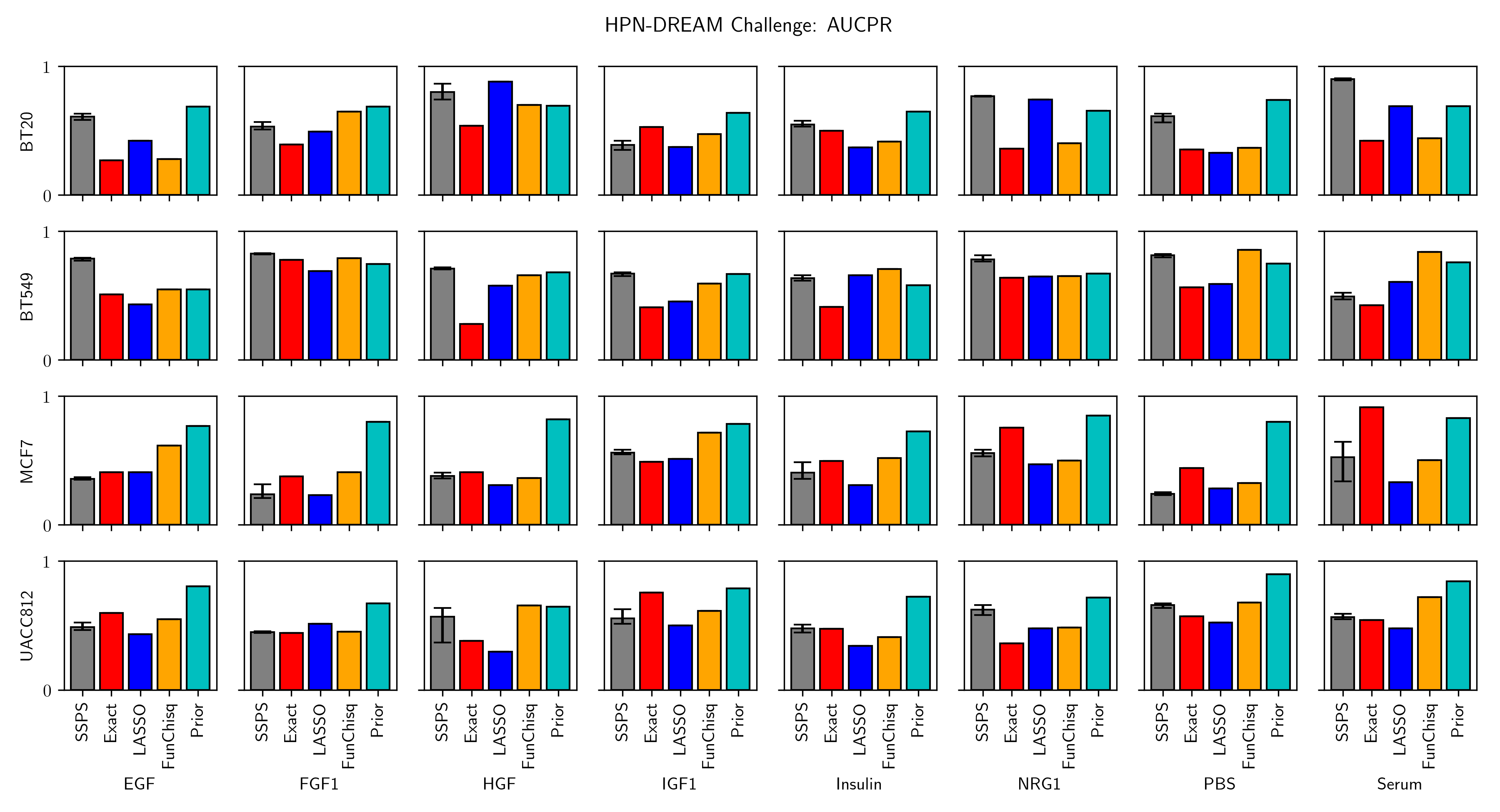}
    \caption{A bar chart similar to Figure \ref{fig:dream-bar} except that it shows AUCPR rather than AUCROC.
             See Figure \ref{fig:dream-bar} for details about the layout.
             }
    \label{fig:aucpr-bar}
\end{figure*}

\paragraph{HPN-DREAM AUCPR.}
For completeness, we complement the AUCROC results of Section \ref{sec:dream-results} with the corresponding AUCPR results.
Figure \ref{fig:aucpr-bar} shows AUCPR in bar charts, with an identical layout to Figure \ref{fig:dream-bar}.

AUCPR leads us to similar conclusions as those from AUCROC.
\Ourmethod{} dominates the exact DBN method in 19 contexts and is dominated in 10.
Both \Ourmethod{} and \Funchisq{} dominate each other in 14 contexts.
However, \Ourmethod{} dominates the prior knowledge in only 9 contexts, and is dominated in 21.
As before, we conclude that \Ourmethod{} attains similar performance to established methods on this task.

\begin{figure*}
    \centering
    \textbf{ROC Curves}\\
    \includegraphics[scale=0.5]{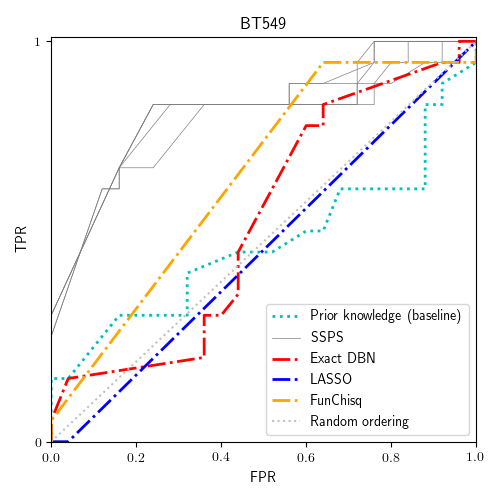}
    \includegraphics[scale=0.5]{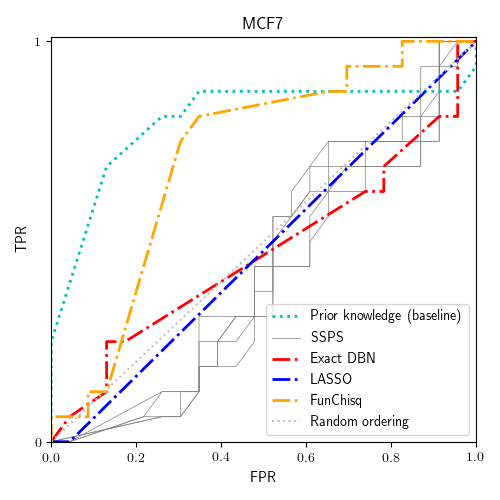}\\
    \vspace{4pt}
    \textbf{PR Curves}\\
    \includegraphics[scale=0.45]{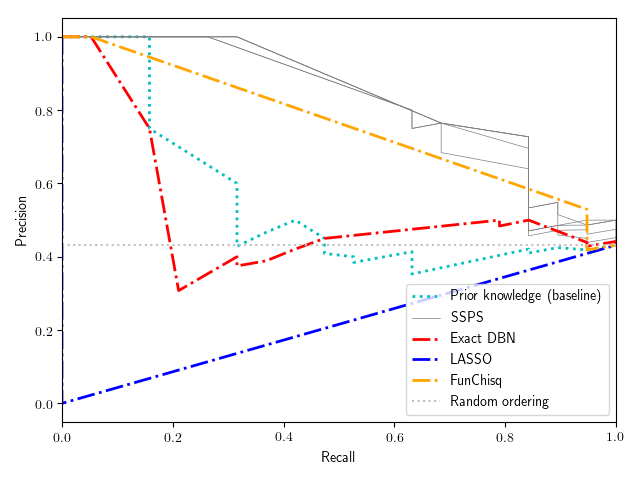}
    \includegraphics[scale=0.45]{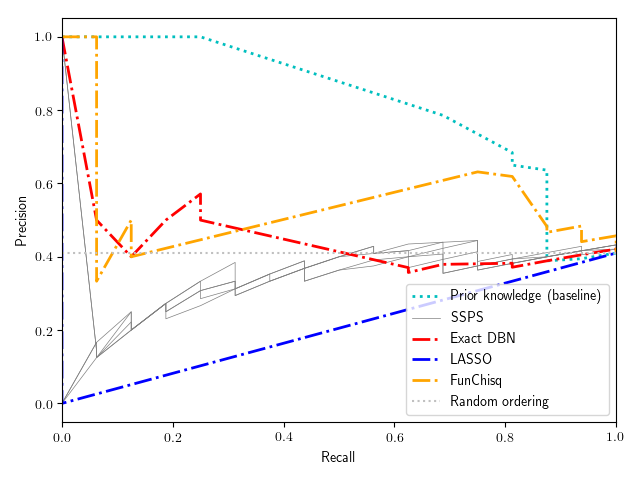}
    \caption{ROC curves (top) and PR curves (bottom) from the HPN-DREAM challenge.
             We show results for two contexts: cell line BT549 (left) and MCF7 (right).
             The stimulus is EGF for both contexts.
             Since SSPS is stochastic, we show all 5 of its curves in each plot.
             The other methods are all deterministic, and therefore only have one curve in each plot.
             }
    \label{fig:curves-dream}
\end{figure*}

\paragraph{ROC and PR curves.}
Figure \ref{fig:curves-dream} shows ROC and PR curves from our HPN-DREAM evaluation.
We focus on two representative contexts: cell lines BT549 and MCF7, with EGF as the stimulus.

The bar charts in Figure \ref{fig:aucpr-bar} tell us that \Ourmethod{} was the top performer in the (BT549, EGF) context.
The ROC and PR curves are consistent with this.
SSPS dominates the other methods in ROC and PR space.
In contrast, \Ourmethod{} was the worst performer in the (MCF7, EGF) context.
The curves show \Ourmethod{} performing worse than random.

The LASSO ROC and PR curves are interesting.
Its ROC curves show nearly random performance.
Its PR curves are straight lines.
Manually inspecting its predictions yields an explanation: 
\primo{} LASSO gives nonzero probability to a very small number of edges;
\secundo{} that small set of edges results in a very small descendant set for mTOR;
\tertio{} that small descendant set is incorrect.

